\documentclass[fleqn,usenatbib]{mnras}

\usepackage[T1]{fontenc}
\usepackage{ae,aecompl}
\usepackage{pdflscape}

\usepackage{graphicx}	% Including figure files
\usepackage{amsmath}	% Advanced maths commands
\usepackage[stable,symbol]{footmisc}
\usepackage{amssymb}	% Extra maths symbols
\hypersetup{draft} % to prevent brain-dead latex behaviour that fails on hyperlinks spanning more than one page
\title[Hydrus 1]{Snake in the Clouds: A new nearby dwarf galaxy in the Magellanic bridge\thanks{This paper includes data gathered with the 6.5 meter Magellan Telescopes located at Las Campanas Observatory, Chile.}}

\author[S. E. Koposov et al.]{
Sergey E. Koposov,$^{1,2}$\thanks{E-mail: skoposov@cmu.edu}
Matthew G. Walker,$^{1}$
Vasily Belokurov,$^{2,3}$ 
Andrew R. Casey,$^{4,5}$
\newauthor 
Alex Geringer-Sameth,$^{6}$
Dougal Mackey,$^{7}$ 
Gary Da Costa,$^{7}$
Denis Erkal$^{8}$,
\newauthor
Prashin Jethwa$^{9}$,
Mario Mateo,$^{10}$,
Edward W. Olszewski$^{11}$ and
John I. Bailey III$^{12}$
\\
$^{1}$McWilliams Center for Cosmology, Carnegie Mellon University, 5000 Forbes Ave, 15213, USA\\
$^{2}$Institute of Astronomy, University of Cambridge, Madingley road, CB3 0HA, UK\\
$^{3}$Center for Computational Astrophysics, Flatiron Institute, 162 5th Avenue, New York, NY 10010, USA\\
$^{4}$School of Physics and Astronomy, Monash University, Clayton 3800, Victoria, Australia\\
$^{5}$Faculty of Information Technology, Monash University, Clayton 3800, Victoria, Australia\\
$^{6}$Astrophysics Group, Physics Department, Imperial College London, Prince Consort Rd, London SW7 2AZ, UK\\
$^{7}$Research School of Astronomy and Astrophysics, Australian National University, Canberra, ACT 2611, Australia\\
$^{8}$Department of Physics, University of Surrey, Guildford, GU2 7XH, UK\\
$^{9}$European Southern Observatory, Karl-Schwarzschild-Str. 2, 85748 Garching, Germany\\
$^{10}$Department of Astronomy, University of Michigan, 311 West Hall, 1085 S University Avenue, Ann Arbor, MI 48109, USA\\
$^{11}$Steward Observatory, The University of Arizona, 933 N. Cherry Avenue., Tucson, AZ 85721, USA\\
$^{12}$Leiden Observatory, Leiden University, Niels Bohrweg 2, 2333 CA Leiden, The Netherlands
}

\date{Accepted XXX. Received YYY; in original form ZZZ}

\pubyear{2018}

\begin{document}
\label{firstpage}
\pagerange{\pageref{firstpage}--\pageref{lastpage}}
\maketitle

\begin{abstract}

We report the discovery of a nearby dwarf galaxy in the constellation
of Hydrus, between the Large and the Small Magellanic Clouds. Hydrus~1
is a mildy elliptical ultra-faint system with luminosity $M_V\sim -4.7$ and 
size $\sim$ 50\,pc, located 28 kpc from the Sun and 24 kpc from the
LMC. From spectroscopy of $\sim 30$ member stars, we measure a velocity
dispersion of 2.7 km\,s$^{-1}$ and find tentative evidence for a
radial velocity gradient consistent with 3\,km\,s$^{-1}$
rotation. Hydrus~1's velocity dispersion indicates that the 
system is dark matter dominated, but its dynamical mass-to-light ratio
M/L $\sim$ 66 is   significantly smaller than typical for
ultra-faint dwarfs at similar luminosity. 
The kinematics and spatial position of Hydrus~1 make it a very 
plausible member of the family of satellites brought into the Milky Way 
by the Magellanic Clouds.
While Hydrus~1's proximity and well-measured 
kinematics make it a promising
target for dark matter annihilation searches, we find no evidence 
for significant gamma-ray emission from Hydrus~1.
The new dwarf is a metal-poor galaxy with a mean metallicity
[Fe/H]=$-2.5$ and [Fe/H] spread of $0.4$ dex, similar to other
systems of similar luminosity. Alpha-abundances of Hyi~1 members
indicate that star-formation was extended, lasting between 0.1 and
1 Gyr, with self-enrichment dominated by SN Ia. The dwarf also hosts a
highly carbon-enhanced extremely metal-poor star with [Fe/H]$\sim$
$-3.2$ and [C/Fe] $\sim$ +3.0.

\end{abstract}


%%%%%%%%%%%%%%%%%%%%%%%%%%%%%%%%%%%%%%%%%%%%%%%%%%

%%%%%%%%%%%%%%%%% BODY OF PAPER %%%%%%%%%%%%%%%%%%

\section{Introduction}

There is no solution to the ``missing satellites'' problem if it is
defined simply as the mismatch between the total count of the Milky
Way dwarf galaxies and the predicted number of dark matter halos capable of
hosting a galaxy, since there is no problem. Once the observed number
of the Galactic satellites is scaled up to correct for the effects of
the detection efficiency and the number of DM sub-halos is scaled down
in accordance with our understanding of the physics of star-formation,
little inconsistency remains \citep[][]{benson02,koposov08,
  tollerud08, koposov09, jethwa18}. However, to make the theory and
the observations match requires that the vast majority of the Galactic
satellites have extremely low luminosities but occupy sub-halos with
significant virial masses. Only in the past 15 years, thanks to the
plethora of imaging surveys such as Sloan Digital Sky Survey
\citep{york00}, Pan-STARRS \citep{kaiser02}, VST ATLAS
\citep{shanks15} and the development of automated search techniques
\citep[e.g.][]{willman05,belokurov07,koposov08,walsh09}, the satellite
census has been expanded to include systems with luminosities of
$L\sim 100~L_\odot$ estimated to inhabit DM halos of at least $\sim
10^7-10^9 M_\odot$ \citep[e.g.][]{simon07,koposov09,jethwa18}. Thus,
the ``missing satellites'' problem is perhaps better viewed as a
dramatic divergence between the sub-halo mass function and the dwarf
luminosity function \citep{klypin99}, with a pile-up of literally
invisible objects with mass-to-light ratios of order $10^5-10^6
M_{\odot} L^{-1}_{\odot}$.

The faintest of these ultra-faint dwarfs are only observable in and
around the Milky Way, thus making the Galaxy a unique place to provide
crucial constraints on theories of Universal structure
formation. However, as a result of the expansion of the discoveries
into the ultra-faint regime, one problem appears to have
emerged. Since most searches have a limiting surface brightness of
$30$ -- $32$ mag arcsec$^{-2}$, many of the recently detected systems
tend to have small physical sizes, often only a few tens of
parsecs. This makes the classification into dwarf galaxies and
globular clusters ambiguous based on the object's size alone
\citep{gilmore07,martin08}. Furthermore, the dynamical tests for the
presence of dark matter in the faint systems are often fraught with
difficulties caused by low numbers of member stars accessible for
spectroscopy \citep{walker09c}, stellar binaries \citep{koposov11} and
instrument systematics \citep{simon07}. As a result, some low
luminosity satellites have oscillated between the two classes
\citep[e.g.][]{belokurov14,laevens14,bonifacio15,weisz16,voggel16}
while the nature of some remains uncertain today
\citep{willman05b,willman11}. To circumnavigate these displeasing
ambiguities, additional diagnostics of ``galaxyness'' have been
offered, including, in particular, one relying on the presence of a
spread of heavy element abundances \citep[see][]{willman12} indicative
of a prolonged star formation activity.

Most recently, our view of the Milky Way satellite population has
become even more curious. In the past 3 years, data from the
various observing campaigns utilizing the Dark Energy Camera - such as
DES, MagLiteS \citep{drlica_wagner16}, and SLAMS \citep{jethwa18b} - have
yielded a stream of new satellite discoveries
\citep{koposov15,bechtol15,kim15a,kim15b,martin15,drlica_wagner15}.
Many of these have been uncovered thanks to the superb quality of
 deep DECam data that has facilitated detection of fainter and more distant
systems. However, many were brought to light simply due to the
expansion of the imaging surveys into the previously barely explored
Southern celestial hemisphere. Remarkably, the bulk of the Southern
discoveries are situated near the Magellanic Clouds \citep{koposov15},
which in hindsight seems obvious. Initially, as the data started to
come in, the satellite over-density signal was barely
significant. However, at present, the satellite positions
\citep{drlica_wagner16,torrealba18} and their kinematics
\citep{koposov15b,walker16} appear to favor the hypothesis in which
the Magellanic Clouds (MCs) have brought a large group of their
smaller companions into the Milky Way \citep{deason15,jethwa16}. At
the same time, thanks to a deeper and more complete view of the
Magellanic area, supplied by the DECam and Gaia data, we have learned
a great deal about the Clouds themselves and their interaction with
each other \citep[see][]{mackey17,belokurov17}.

Around the Clouds, the relatively-unexplored portion of the sky that has
the greatest potential to inform our understanding of both the satellites
of the Magellanic satellites and the Clouds' life as a pair is the
area directly between the LMC and the SMC. This paper presents results
from a continuation of the survey of the Magellanic bridge
region, first presented by \citet{mackey17}. While this survey was
motivated primarily by the study of the stellar tidal debris field
around the Magellanic Clouds \citep[see e.g.][]{mackey16,belokurov16},
it also provides a great resource to trawl for new MC
satellites. Here we report the discovery and a detailed photometric,
spectroscopic and chemical study of exactly that: a new dwarf galaxy
in the bridge region of the Magellanic Clouds.  The paper is
structured as follows. In Section~\ref{sec:data} we describe the
photometric data from Dark Energy Camera. In
Section~\ref{sec:sat_properties} we discuss the satellite properties
that we infer from the photometric data. In
Section~\ref{sec:spectroscopy} we discuss the spectroscopic data and
modeling of a new satellite. Section~\ref{sec:discussion} provides a
discussion of all the chemical, dynamical and morphological 
properties of the new system, possible connections to the Magellanic Clouds 
and prospects for dark matter annihilation searches. The Section~\ref{sec:conclusions}
concludes the paper. The properties of the
stellar debris field around the Magellanic Clouds are discussed in a
companion paper by Mackey et al (2018).

\section{Photometric Data}
\label{sec:data}
This work utilizes the photometric data from two DECam programs,
2016A-0618 and 2017B-0906, focusing on the region between the
Magellanic Clouds. The first results based on the 2016A-0618 imaging
have been presented in \citet{mackey17}. The DECam is a large (570
mega-pixel) mosaic camera boasting 62 science CCDs, giving a combined
field-of-view of 3 square degrees \citep{flaugher15}. The camera is
installed on the 4-m Blanco Telescope located at Cerro-Tololo
Inter-American Observatory (CTIO) in Chile. The camera is equipped
with $ugrizY$ filters and has been used for the last 6 years to
conduct the Dark Energy Survey \citep{des05,abbott18} as well as a
number of individual programs and mini-surveys
\citep{drlica_wagner16,schlafly18}.

The data for the program 2017B-0906 - the primary data source for this
paper - were acquired on 8 and 9 October 2017. We observed the area 
covering the side of the Magellanic bridge region closest to the Small
Magellanic Cloud. Images were taken in $g$ and $r$ filters
with exposure times ranging from 70 to 225 s in the g band and from 
60 to 110 s in the r band, where the variation was due to changing
lunar illumination during the night. On average, in the survey, each
sky location was covered by 3 individual exposures and the total
observed area is approximately 140 square degrees.

The initial data processing was carried out using the community
pipeline \citep{valdes14} which provided us with calibrated single
exposure images that were corrected for bias, flat-field, cross-talk,
nonlinearity as well as being WCS-aligned and remapped. The subsequent image
processing was very similar to that described by \citet{mackey17}. We
first created image stacks in $g$ and $r$ bands from all of the
available data using the SWARP software \citep{bertin10}. The sources
extracted from each CCD frame were used to construct the Point Spread
Function (PSF) by running PSFex \citep{bertin11}. Then we used a
custom version of the SExtractor code \citep{bertin96} in double image
mode on stacked and individual images to obtain uncalibrated forced
photometry PSF fluxes from individual images. The fluxes were then
photometrically calibrated and averaged for sources with more than one
observation. The lists of detected $g$ and $r$-band sources were
cross-matched to produce the final catalogue with a magnitude limit of $\sim 23$ in both $g$ and $r$.

The photometric calibration was done using the APASS DR9 catalogue
\citep{henden12}. Because APASS has significant spatially dependent
systematic errors reaching up to $0.05$-$0.1$ mag, we first rectified
the APASS magnitudes using Gaia \citep{gaia16,brown16} and 2MASS
\citep{skrutskie06} data. To perform the APASS correction, we assumed
that the APASS $g$ and $r$ magnitude for stars with $0<J-K_s<0.75$
could be estimated from the Gaia $G$ band measurement and the
polynomial function of the 2MASS $J-K_s$ colour. This allowed us to
compute the $g$ and $r$ magnitude HEALPIX \citep{gorski05} correction
map for APASS with resolution of $N_{side}=128$ (corresponding to
$\sim 0.5\degr\times 0.5\degr$ degrees).  Finally we brought the
magnitudes from the APASS photometric system into the DES DR1 system
\citep{abbott18} using simple color-terms.  The recalibrated APASS
magnitudes together with the information from the overlapping DECam
exposures were then used to establish zero-points for each DECAM
exposure as well as the CCD-specific zero-point offsets (similarly to
\citet{koposov15}). A comparison of the final calibrated catalogue
with the small patches of the DES DR1 that overlap with our footprint
reveals that the remaining systematic errors in the catalogue are at a
level $\lesssim 0.03$ mag.

Throughout the paper we use extinction corrected magnitudes based on
the \citet{schlegel98} extinction model with the coefficients from
\citet{schlafly11}. To separate stars from galaxies in our catalogue
we use the criterion based on the {\tt SPREAD\_MODEL} parameter
measured by Sextractor in the $r$ band: $|SPREAD\_MODEL|<0.003$
\citep{desai12}.

\begin{figure}

	\includegraphics[width=\columnwidth]{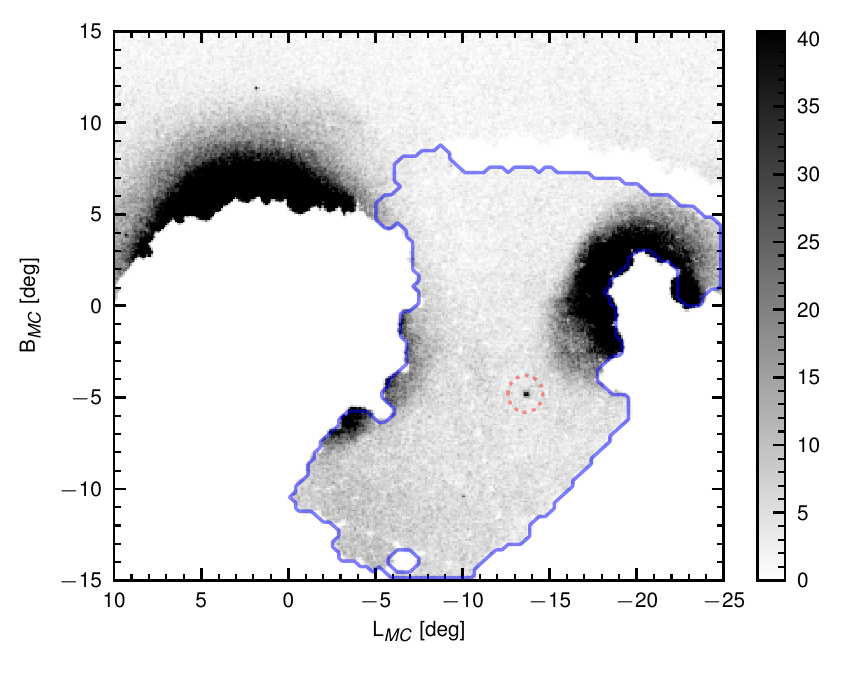}
    \caption{The density of MSTO stars ($20.5<r<22$ and $0<g-r<0.3$ in the Magellanic Clouds region. We include data from our new DECam observing program, data from \citet{mackey17} and data from the  Dark Energy Survey DR1. The data from our Magellanic bridge observations is surrounded by blue contour. The location of newly discovered dwarf galaxy in between Large Magellanic Cloud (on the left) and Small Magellanic Cloud (on the right) is marked by a red dashed circle. The coordinate system used for the plot is a rotated spherical coordinate system such that the equator is going through both LMC and SMC centers. The pole of this coordinate system is  $\alpha_{pole},\delta_{pole}=39.61\degr,15.49\degr$ and the longitude zeropoint is at the location of LMC.}
    \label{fig:lmc_map}
\end{figure}

In order to ensure a uniform calibration for both of our datasets,
2016A-0618 and 2017B-0906, we reprocessed the data from 2016 together
with data from 2017 to obtain a continuous coverage of $\sim$ 380
square degrees across the inter-Cloud region. Figure~\ref{fig:lmc_map}
shows the density distribution of distant $20.5<r<22$ $0<g-r<0.3$ main
sequence turn-off (MSTO) stars selected from our two Magellanic Cloud
programs as well as the data from the Dark Energy Survey Data Release 1. 
The combined footprint of our programs is traced by a blue contour showing 
the targeted region of sky. We remark that the data do
not show any discontinuity either between our 2016 and 2017 data or
between our program and the DES survey, thus confirming the absence of
significant systematic offsets in our photometry.

A stand-alone feature immediately apparent in the stellar density map
in Figure~\ref{fig:lmc_map} is an overdensity at $L,B=(-13,-5)$.  This
overdensity in the Hydrus\footnote{The Hydrus constellation is different from
  Hydra. Hydrus is a mythical creature usually represented by a male
  water snake.} constellation is the subject of this paper. In what
follows we classify the satellite as a dwarf galaxy (see
Section~\ref{sec:spectroscopy}) and therefore it is referred to as
Hydrus~1 or Hyi~1 in the rest of the text.

\section{Satellite properties}
\label{sec:sat_properties}

\subsection{Stellar population} 
\label{sec:stellar_pop}

\begin{figure}
	\includegraphics[width=\columnwidth]{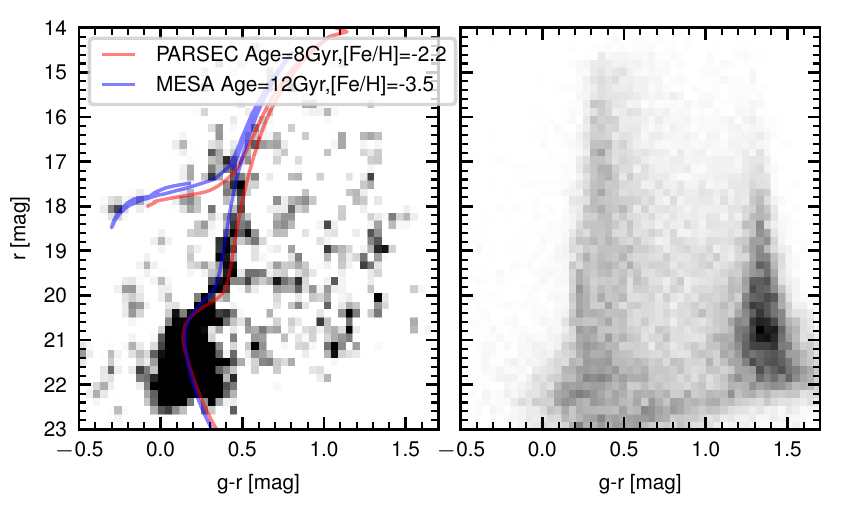}
    \caption{{\it Left panel:} Background subtracted $g$-$r$, $r$
      Hess diagram within 10\arcmin\ of Hydrus~1. Stars within an
      annulus with radii of 20\arcmin\ and 60\arcmin\ were used for
      the background. Two theoretical isochrones shifted to a distance
      of 31\,kpc from MESA and PARSEC libraries are overplotted in
      blue and red respectively. {\it Right panel:} Hess diagram
      of background stars around Hydrus~1 (in 20\arcmin,
      60\arcmin\ annulus).}
    \label{fig:cmd}
\end{figure}

We start by analyzing the color-magnitude diagram (CMD) of the
overdensity circled in red in Figure~\ref{fig:lmc_map}. The left panel
of Figure~\ref{fig:cmd} shows the background-subtracted Hess diagram
of the central 10\arcmin\ of the object. A clear main sequence (MS)
together with a part of the red giant branch are both readily visible, 
confirming that the object is a {\it bona-fide} satellite. The Hess
diagram of the overdensity is also clearly different from the
color-magnitude distribution of foreground stars around it shown in
the right panel of the Figure. To measure the distance to the
satellite and constrain its stellar population properties, we perform
a fit to the Hess diagram of the system within the inner 10\arcmin\ by
using two different sets of theoretical isochrones. More precisely, we
use both PARSEC isochrones \citep{bressan12,marigo17}\footnote{We use
  the PARSEC v1.2S,COLIBRI PR16 isochrones downloaded from
  \url{http://stev.oapd.inaf.it/cgi-bin/cmd} in February 2018} and
MESA isochrones \citet{dotter16,choi16,paxton11}. The isochrones are
populated according to the Chabrier Initial Mass Function (IMF)
\citep{chabrier03} with an additional magnitude-dependent photometric
scatter based on the uncertainties in the data. We then evaluate the
likelihood of the observed CMD over a grid of ages, metallicities and
heliocentric distances. The resulting best-fit MESA and PARSEC
isochrones are overplotted in the left panel of the
Figure~\ref{fig:cmd}.

The best parameters for the MESA isochrone are: age of $12$\,Gyr, iron
abundance of ${\rm [Fe/H]}=-3.5$ and distance of $31$\,kpc. On the
other hand, for the PARSEC isochrone we obtain $8$\,Gyr, ${\rm [Fe/H]}=-2.2$ and
distance of $\sim 31$\,kpc. We find that MESA isochrones fit the
observed Hess diagram somewhat better, possibly because the data are pushing the PARSEC library to its lower limit of [Fe/H]=$-2.2$. Due to
large systematic differences in isochrone solutions we do not try to
assess the uncertainties in the age and metallicites inferred, however
the formal uncertainty on the heliocentric distance for each isochrone
set is around $2$\,kpc. While CMD fitting is an acceptable method of
deriving distances it is also known to suffer from inaccuracies
associated with the age-metallicity-distance degeneracies as well as
biases associated with the isochrone libraries used. These issues can
be mitigated by using horizontal branch stars (both pulsating and not)
which can offer better accuracy in distance in most cases. In fact in
Figure~\ref{fig:cmd} we can see $\sim$ 4 stars at r $\sim 18$ and
g-r$\lesssim 0$ that are likely Hydrus's Blue Horizontal Branch (BHB)
stars (some of these were later spectroscopically confirmed; see
Section \ref{sec:spectroscopy}). This group of four likely BHB stars can
be used to measure the distance by applying the fiducial relation for
the horizontal branch from \citet{deason11}. As that relation was
defined for the SDSS photometric system, we recalibrate it into the
DES system.  The resulting equation that we adopt is:
 $$M_g = 0.444 + 0.106\, c + 0.704\,c^2 + 1.126\,c^3+0.794\,c^4 $$
where $c = (g-r)/0.3$.  With four BHB stars we derive the distance
modulus $m-M=17.20\pm 0.04$ or distance of $D=27.6\pm 0.5$\, kpc,
which is roughly consistent with the isochrone fit result. The
expected systematic uncertainty on the BHB measurement should be of
the order of 10\% \citep{fermani13}.

To validate the distance measurement we have employed the Magellanic
RR Lyrae catalog from the OGLE-IV project \citep{udalski15} which
covers the location of Hydrus~1.  Somewhat suprisingly, within twice
the half-light radius of Hyi~1 we identified two RRab Lyrae stars,
OGLE-SMC-RRLYR-6316 and OGLE-SMC-RRLYR-6325, with distance moduli of 17.22
and 17.19 respectively\footnote{To compute distances to the RR Lyrae
  we use the period-luminosity relations from \citet{catelan04},
  metallicity phase relation from \citet{smolec05} as prescribed by
  \citet{pietrukowicz15}}. With an average density of 0.12 stars\,deg$^{-2}$ for RR Lyrae in the distance modulus range $17<m-M<17.4$, the probability of having two Galactic RR Lyrae within twice the half light radius of Hyi~1 is $P \approx 2 \times 10^{-4}$.  This suggests that both RR lyrae are most likely genuine members of Hyi~1. The RR Lyrae fully corroborate the distance estimate of $m-M=17.2$ that we obtained based on BHB magnitudes.

With period and V-band amplitude combination of (0.67d, 0.8) and
(0.73d, 0.56), the two RR Lyrae in Hyi~1 are of Oosterhoff type
II. This is in perfect agreement with the recent discovery that the
fraction of Oo I/Oo II types depends strongly on the stellar mass of
the host \citep[see e.g.][]{fiorentino15, belokurov_rrl}, with higher
luminosity objects (e.g. massive dwarf galaxies) containing a larger
fraction of Oo I RRab stars. Extended down to the typical luminosities
of UFDs, this trend predicts that the RR Lyrae contents of objects
like Hyi~1 should be almost entirely composed of Oo II type pulsators.

\subsection{Spatial distribution}
\label{sec:spatial_distrib}

\begin{figure}
	\includegraphics[width=\columnwidth]{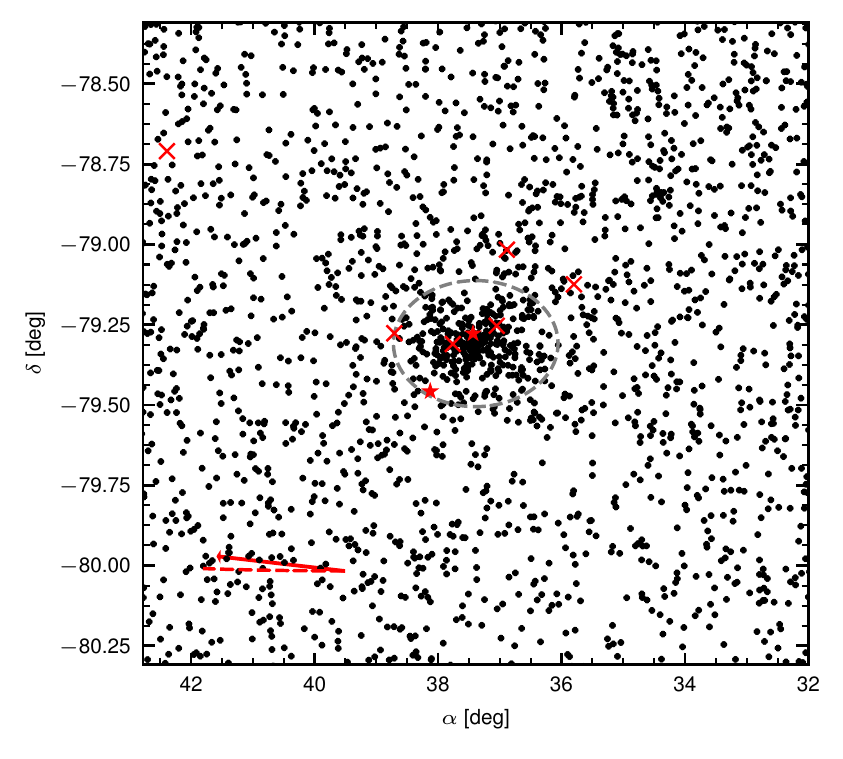}
    \caption{The distribution of MSTO ($0<g-r<0.3$, $20.5<r<22$) stars in the vicinity of  the Hydrus~1 satellite. The dashed grey ellipse shows the best-fit shape of the satellite where the major axis of the ellipse is twice the half-light radius. The arrow in the left corner shows the direction of the LMC's proper motion and the dashed line is parallel to the line connecting LMC and SMC. Red crosses show the locations of BHB candidate stars with distance modulus within 0.1 magnitude of Hydrus~1, while red star symbols identify the locations of RR Lyrae stars with distance modulus within 0.1 magnitudes of Hydrus~1. We note that both the RR Lyrae and BHB stars show clear concentrations near the center of the object. }
    \label{fig:zoom}
\end{figure}

This section looks in detail at the spatial distribution of stars
inside the Hydrus~1 satellite. Figure~\ref{fig:zoom} gives the
positions of the MSTO stars with $0<g-r<0.3$, $20.5<r<22$ in $1$
square degree region around the center of the satellite. We clearly
see a high-contrast over-density with some hint of elongation
approximately along the right ascension direction. In the Figure we
also overplot the locations of the BHB stars and the RR-lyrae discussed
in the previous section, thus confirming that they are clustered near the center of Hydrus~1.  To describe the morphological
properties of Hyi~1, we model the stellar distribution with a
combination of uniform background and flattened exponential profile
\citep[similar to ][]{koposov15}. We adopt standard priors on
the model parameters: a Jeffrey's prior on the exponential scale length,
uniform $U(0,1)$ priors on the object ellipticity and the fraction of
stars belonging to Hydrus~1, a $U(0,180)$ prior on the position
angle and an improper uniform prior on the position of the center of the
object. We then sample the posterior using a Markov Chain Monte-Carlo
(MCMC) technique, specifically with the ensemble sampler
\citep{goodman10} as implemented by \citet{foreman_mackey13}. The
summary of our measurements is provided in
Table~\ref{tab:measurements}. Here and throughout the rest of the
paper, when we report the measurements and their associated
uncertainties we use the mean and standard deviation in the case of
symmetric Gaussian-like posteriors and the median and 15.8 and 84.2
percentiles for asymmetric posteriors.  In the Table we report both
the major-axis half-light radius (i.e $1.67\,h$ where $h$ is the
exponential scale-length along the major axis) as well as the
circularized half-light radius (i.e. multiplied by $\sqrt{1-e}$).

\begin{figure}
	\includegraphics[width=\columnwidth]{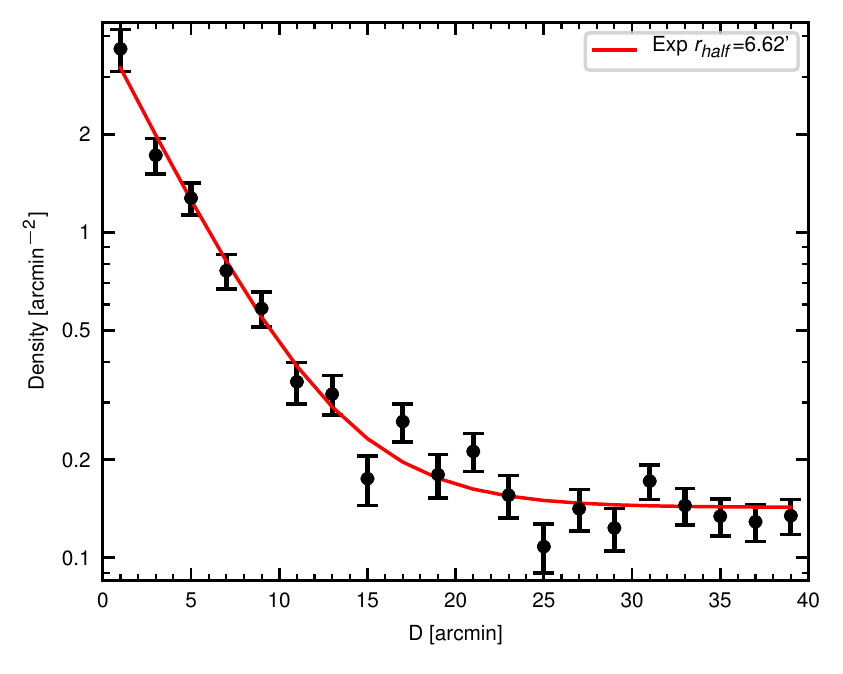}
    \caption{
    Azimuthally averaged density profile of Hydrus~1 turn-off stars with $0<g-r<0.3$, $20.5<r<22.25$. The red curve shows our best fit exponential model.}
    \label{fig:dens_prof}
\end{figure}

The angular half-light radius of the satellite is $\sim 6.6\arcmin$
while its physical size, assuming a distance of 28\,kpc, is $\sim$
50\,pc.  These values suggest that Hyi~1 is a dwarf galaxy rather than a globular cluster \citep[see e.g. Figure 6 of][]{torrealba16}. We observe that
the ellipticity of the system is significantly different from zero:
$e=1-\frac{b}{a}\sim0.2$, confirming its slightly elongated visual
appearance in Figure~\ref{fig:zoom}. It is worth noting that the
elongation of the system is approximately aligned with the Magellanic
bridge and the proper motion of the Magellanic Clouds (shown as a red
dashed line segment and a red arrow, respectively, in
Fig.~\ref{fig:zoom}). For now, we refrain from speculation about whether
this is a mere coincidence or a tell-tale sign of a connection between 
Hydrus~1 and the Magellanic Clouds.  To verify whether the exponential model adopted for Hyi~1's
density profile provides an accurate description of the data,
Figure~\ref{fig:dens_prof} displays the azimuthally averaged 1-D
density profile of the satellite together with the best-fit
exponential model. Reassuringly, the model reproduces the observed
profile well at all radii.

Equipped with a color-magnitude model and a spatial model of the
satellite we can proceed to measure the absolute luminosity of the
system. We first compute the number of stars in Hydrus~1 that are near our best-fit isochrone and brighter than $r=21.75$.  This is approximately $1$
magnitude above the limit of our photometric data and therefore should
give us an almost complete sample. Above this limit Hyi~1 has an
estimated 387 $\pm$ 28 stars, which, assuming the best-fit MESA
isochrone at the distance from Section~\ref{sec:stellar_pop} and a
Chabrier IMF, gives a total stellar mass of $\sim$ 6000 M$_\odot$
and an expected V-band absolute luminosity of $-4.74 \pm
0.08$\footnote{We remark that the total luminosity that we compute is
  the expected luminosity given the total stellar mass and mass
  function, rather than the estimate obtained by summing luminosities
  of individual stars, which tends to suffer from large fluctuations due
  to poorly populated red giant branch.}.  We also checked that using
the PARSEC isochrone yields the same absolute magnitude within the
uncertainty.

\begin{table*}
\begin{center}
\begin{tabular}{c|c|c|c}
\hline
Parameter & Description & Value & Unit \\
\hline
$\alpha_{cen}$ &  Right ascension of the center & $37.389 \pm 0.031$& degree\\
$\delta_{cen}$ &  Declination the center &$-79.3089\pm 0.0045$&degree\\
$r_{half,maj}$ &  Major axis half light radius & $7.42^{+0.62}_{-0.54}$& arcmin\\
$r_{half,maj}\sqrt{1-e}$ & Circularised half-light radius & $6.64^{+0.46}_{-0.43}$ & arcmin\\
$r_{half,maj}\sqrt{1-e}$ & Circularised half-light radius & $53.3\pm3.6$& pc \\
$1-b/a$ & Ellipticity &$ 0.21^{+0.15}_{-0.07}$\\
PA & Positional angle of the major axis of stellar distribution & $97\pm 14$& degree\\
m-M & Distance modulus & $17.20 \pm 0.04$ \\
Distance & Heliocentric distance & $27.6 \pm 0.5$ & kpc \\
$M_V$ & Absolute Magnitude & $-4.71 \pm 0.08$ & mag\\
$\sigma_{los}$ & Line of sight velocity dispersion &  $2.69_{-0.43}^{+0.51}$ & km\,s$^{-1}$\\
$V_{hel}$ &  Mean Heliocentric radial velocity & $80.4\pm 0.6$ & km\,s$^{-1}$  \\
$V_{GSR}$ & Mean radial velocity in the Galactic rest frame\footnotemark  & $-94.1\pm 0.6$ & km\,s$^{-1}$  \\
M(r<r$_{1/2}$)  & Total dynamical mass inside half-light radius\footnotemark & $2.23_{-0.66}^{+0.95} 10^5 $& $M_\odot$\\
M/L(r<r$_{1/2}$) & Mass-to-light ratio inside the half-light radius & $66_{-20}^{29}$ & M$_\odot$/L$_{\odot}$ \\
$\log_{10}$ M/L(r<r$_{1/2}/(M_\odot/L_{\odot})$) & Logarithm of mass-to-light ratio inside the half-light radius & $1.82\pm0.16$ & dex  \\
$\log_{10}[J(0.5\degr)/ (\mathrm{GeV}^2\mathrm{cm}^{-5})] $ & Logarithm of the $J$ factor & $18.33_{-0.34}^{+0.38}$ &  dex \\
$\frac{dV_{hel}}{d\alpha \cos{\delta_{cen}}}$ & Radial velocity gradient along right ascension & $24.4\pm 9.6$ & km\,s$^{-1}$\,deg$^{-1}$ \\
$\frac{dV_{hel}}{d\delta}$& Radial velocity gradient along declination & $4.1\pm 9.2$ & km\,s$^{-1}$\,deg$^{-1}$\\
$<[Fe/H]>$ & Mean spectroscopic metallicity & $-2.52\pm 0.09$  & dex \\
$\sigma_{[Fe/H]}$ & Spectroscopic metallicity scatter & $0.41\pm 0.08$  & dex \\

\hline
\end{tabular}
\caption{Key photometric and kinematic properties of the Hydrus~1 dwarf galaxy. Reported measurements and uncertainties are either means and standard deviations for symmetric posteriors, or medians and 15.8 and 84.2 percentiles in the case of asymmetric posteriors.}
\label{tab:measurements}
\end{center}
\end{table*}

\section{Spectroscopy}
\label{sec:spectroscopy}

In order better understand the nature of Hydrus~1, we acquired spectra of its brightest member stars.  

\subsection{Target selection}
We selected potential members of the Hyi~1 system as those that lie spatially near the center of Hyi~1 and photometrically near the best-fitting isochrone measured in Section~\ref{sec:stellar_pop}. The leftmost panel of Figure~\ref{fig:spec_plot} shows the spatial distribution of all 235 spectroscopic targets (grey and black points), and the second panel from the left shows the extinction corrected color-magnitude distribution of those same stars.

%FOOTNOTES FOR THE TABLE, I HAD TO FIND LOCATION BY HAND
\addtocounter{footnote}{-1} % I HATE LATEX!!!
\footnotetext{Computed assuming the $V_{LSR}=232.8$\, km\,s$^{-1}$ from \citet{mcmillan17} and peculiar solar velocity from \citet{schonrich10}}
\addtocounter{footnote}{1}
\footnotetext{Obtained using the formula of \citet{walker09b}}

\subsection{Spectroscopic data}
\label{sec:spec_data}
On 12-13 Feb 2018, we used M2FS, a multi-object fiber spectrograph at the Magellan Clay 6.5\,m telescope, to acquire spectra for the selected targets. We used the same instrument configuration as in our previous M2FS observations of ultra-faint dwarf galaxy candidates \citep{walker15b,walker16}---specifically, we observed the spectral region 5130 - 5190 \AA\ at resolution $\mathcal{R}\sim 18,000$.  We obtained $3\times 2700$-second exposures (the middle exposure lost $\sim 300$ seconds due to technical problems) during conditions that were clear, with good seeing but high airmass ($\sim 1.7 - 2.0$).  For calibration purposes, we also obtained high-S/N exposures of the solar spectrum during evening twilight, as well as Th-Ar arc-lamp images immediately before and after the science exposures.  

We processed and fit stellar-atmospheric models to all spectra following procedures identical to those described by \citet{walker15b,walker16}.  For each star, we obtain simultaneous estimates of line-of-sight velocity, effective temperature, surface gravity and metallicity.
For all the parameters we adopt uninformative flat priors with the exception of the effective temperature, for which we use a log-normal prior based on the $g-r$ color and spectroscopic metallicity. Our prior takes the form of $\log T_{eff} \sim {\mathcal N}(S_1(g-r) + A\,\exp{(B\,[Fe/H])},S_2(g-r))$ where A, B are fitted constants and $S_1(g-r)$ and $S2(g-r)$ are cubic splines with 20 equidistant knots in the $-0.3<g-r<1.2$ range. We calibrate this prior using the cross-match of the Dark Energy Survey DR1 photometric catalogue with the SDSS DR9 spectroscopic catalogue, where the $g-r$ colors were taken from DES and metallicities and effective temperatures from measurements by SDSS Spectroscopic Parameter Pipeline \citep{lee08a,lee08b,allende_prieto08}. 
The median parameter uncertainties for the spectra with signal to noise larger than three are $\sigma_{v}=0.82$\,km s$^{-1}$,
$\sigma_{\rm Teff}=80$\,K,
$\sigma_{\rm logg}=0.19$\,dex
and $\sigma_{\rm Fe/H}=0.13$\,dex. The velocity uncertainties include a systematic floor, added in quadrature, of 0.2\,km\,s$^{-1}$, as measured from repeated observations of twilight spectra.  

\begin{table*}
\tiny
\begin{tabular}{cccccccccccccccc}
\hline
ID & $\alpha$ & $\delta$ & $S/N$ & $V_{hel}$ & $\sigma_{V}$ & T$_{eff}$ & $\sigma_{T,eff}$ & $\log g$ & $\sigma_{\log g}$ & [Fe/H] & $\sigma_{[Fe/H]}$ & $\ln \frac{P_{m}}{P_{nm}}$ & g & r \\
 & degree & degree &  & km\,$s^{-1}$ & km\,$s^{-1}$ & K & K &  &  & dex & dex &  & mag & mag \\
\hline
2 & 37.69 & -79.36 & 15.4 & 53.5 & 0.5 & 5723 & 64 & 4.53 & 0.12 & -0.74 & 0.06 & $-64.1_{-24.7}^{+17.3}$ & 18.45 & 17.92 \\
3 & 37.70658 & -79.35572 & 21.7 & 5.0 & 0.5 & 5353 & 35 & 4.49 & 0.08 & -0.7 & 0.04 & $-415.8_{-169.4}^{+119.8}$ & 17.86 & 17.25 \\
4 & 37.74029 & -79.35434 & 9.7 & 83.1 & 1.7 & 5521 & 82 & 1.4 & 0.46 & -2.34 & 0.15 & $7.0_{-0.9}^{+1.0}$ & 19.03 & 18.44 \\
5 & 37.694 & -79.34884 & 6.0 & 211.9 & 1.1 & 5904 & 114 & 1.16 & 0.84 & -1.75 & 0.2 & $-1168.4_{-488.1}^{+343.8}$ & 19.74 & 19.24 \\
6 & 37.72267 & -79.33451 & 18.1 & -6.1 & 0.5 & 5437 & 33 & 4.81 & 0.1 & -0.54 & 0.04 & $-545.6_{-224.0}^{+157.1}$ & 18.21 & 17.63 \\
7 & 37.69496 & -79.32996 & 2.1 & 243.7 & 2.9 & 6269 & 240 & 3.35 & 0.86 & -0.9 & 0.45 & $-1812.6_{-755.8}^{+533.2}$ & 20.89 & 20.45 \\
10 & 37.79812 & -79.40667 & 4.7 & 83.9 & 1.5 & 5876 & 111 & 4.52 & 0.29 & -1.58 & 0.19 & $1.3_{-3.0}^{+1.6}$ & 19.81 & 19.29 \\
11 & 37.83383 & -79.38515 & 10.0 & 6.3 & 0.6 & 5792 & 91 & 3.18 & 0.19 & -1.27 & 0.11 & $-396.6_{-164.5}^{+116.7}$ & 18.84 & 18.33 \\
12 & 37.75471 & -79.349 & 3.7 & 187.4 & 1.5 & 5502 & 111 & 2.84 & 0.97 & -1.88 & 0.27 & $-765.2_{-321.1}^{+227.0}$ & 20.27 & 19.66 \\
14 & 37.81337 & -79.32914 & 2.2 & 85.7 & 6.5 & 6387 & 230 & 2.86 & 1.1 & -1.0 & 0.47 & $-11.1_{-9.9}^{+8.8}$ & 20.81 & 20.4 \\
\hline
\end{tabular}
\caption{Spectroscopic measurements for individual stars with either S/N$>$3 or radial velocity error less than $10$\,km\,s$^{-1}$ in the Hydrus~1 field. We include the internal ID of each star, right ascension, declination, signal to noise of the spectra, heliocentric radial velocity, effective temperature, surface gravity together with their uncertainties. The log-odds ratios of being a Hydrus~1 member star vs being a foreground contaminant from the chemo-dynamical model described in Sec.~\ref{sec:spec_modeling} and the $g$, $r$-band photometry from DECam observations are also included. Only a subset of stars is shown in the printed version of the table, the full version of the table is available online.}
\label{tab:spec_stars}
\end{table*}
%\end{landscape}

\begin{figure*}
	\includegraphics{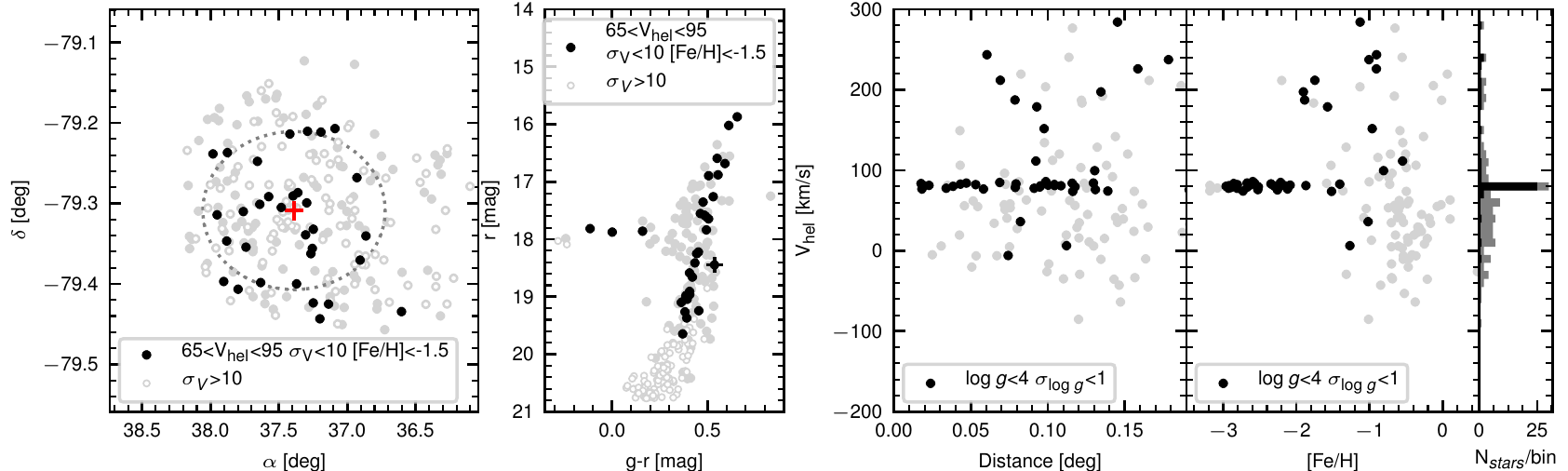}
    \caption{{\it Left panel:} Spatial distribution of spectroscopically targeted stars. The center of Hyi~1 is shown a by red cross.  Darker circles indicate likely members, based on velocity and metallicity, while stars without a good velocity measurement are shown by empty gray circles.  Filled grey circles show all other stars. The dotted line shows the half-light ellipse. {\it Second panel:} Color-magnitude diagram of spectroscopically targeted members. The meaning of the symbols is the same as in the left panel. We note that several horizontal branch stars are marked as likely members; however, a few stars at the very edge of the horizontal branch avoid this distinction due to high velocity uncertainties caused by lack of lines.  A cross-hair marker indicates the carbon-enhanced metal poor star (Star 75) that is a member of Hydrus~1. {\it Third panel:} Heliocentric line-of-sight velocity vs angular distance from Hyi~1's center, for all stars with good velocity measurements (errors $\sigma_v<10$\,km\,s$^{-1}$). Stars classified confidently as giants according to their spectra are shown by darker markers. {\it Fourth panel:} Heliocentric velocity vs spectroscopic metallicity for stars with good velocity measurements. The symbols used are the same as in the previous panel. {\it Rightmost panel:} Velocity histogram for the giant-only subsample (black) and all the stars with good velocity measurements (grey). }
    \label{fig:spec_plot}
\end{figure*}

Of the 235 spectroscopically observed stars, 117 have signal-to-noise larger than 3 and 136 stars have radial velocity (RV) uncertainty less than 10\,km\,s$^{-1}$. As large RV uncertainties of 10\,km\,$s^{-1}$ and more occur with our high-resolution spectra only when the spectrum lacks discernable spectral features and/or has too low signal to noise, we exclude all the stars with $\sigma_v>10$\,km\,s$^{-1}$ from further analysis. We show these stars in the second panel of Figure~\ref{fig:spec_plot} by empty circles. As expected,  most of these stars are faint and have magnitudes r$\gtrsim 19.5 $, with the exception of two blue/hot stars that are likely BHBs that lack strong features within the observed wavelength range. 

Table~\ref{tab:spec_stars} presents spectroscopic parameter measurements for all the stars that either have signal-to-noise larger than 3 or have radial velocity uncertainty less than 10 km\,s$^{-1}$. We list measurements of the effective temperature, surface gravity, metallicity and heliocentric line-of-sight velocitiy.  We also provide in the table the $g$, $r$ magnitudes and the log-odds of being Hydrus~1 member star, based on the model described in the next Section. 

The right three panels of Figure~\ref{fig:spec_plot} summarize our spectroscopic results. Only stars with RV uncertainty smaller than 10\,km\,s$^{-1}$ are shown. The third panel of the Figure shows the heliocentric velocity of observed stars vs distance from Hydrus's center, and the fourth panel shows line-of-sight velocity vs spectroscopic metallicity. The fifth panel shows the 1-D distribution of velocities. These panels clearly confirm that Hydrus~1 is a real stellar system. A large concentration of stars belonging to Hyi~1 is prominent at heliocentric velocities of $\sim$80\, km\,s$^{-1}$ and low metallicities $[{\rm Fe/H}]<-2$. This group of stars is kinematically cold and is noticeably more metal-poor than other stars in the field that likely belong to the halo and/or MW disk. In total our sample seems to contain around $30$ Hydrus~1 member stars. In the three rightmost panels of the Figure we also highlight the stars that are confidently classified as giants/sub-giants, i.e. $\log g<4$, $\sigma_{\log g}<1$ (shown in black rather than grey). This particular subsample of stars offers a particularly clear view of the Hydrus~1 RV peak at 80\,km\, s$^{-1}$, with little contamination at different heliocentric velocities. This is expected given that stars belonging to Hydrus~1 should be mostly giant branch or horizontal branch stars and the contamination should mostly consist of thick disk dwarf stars and, to a much lesser extent, stellar halo giant stars at distances of 30\,kpc.

\subsection{Chemo-Dynamical Modeling}
\label{sec:spec_modeling}

As a next step we proceed to modeling of the observed kinematic and chemical properties of Hyi~1 stars. To describe the distribution of velocities and chemical abundances in the data, we 
adopt a standard mixture model with different components representing Hyi~1, MW halo and disk. We also utilise the fact that the sample of stars classified as giants is much cleaner and should not be contaminated by thin and/or thick disk stars. Therefore we split our star sample into two subsamples: stars confidently classified as giants ($\log g< 4$, $\sigma_{\log g}<1$) and dwarfs (the rest). 

A brief summary of the model is given below, where we provide the distribution of velocity $v$ and metallicity $[Fe/H]$ conditional on the object type $T$, a latent categorical variable describing whether a given star is a halo star ($T={\bf h}$), disk star ($T={\bf d}$) or Hydrus~1 (object) star ($T={\bf o}$). For all the components: halo, disk and Hydrus~1 we assume that [Fe/H] abundances and velocities are independently Gaussian distributed. We also assume that radial velocity of Hydrus~1  stars could have a linear gradient on the sky, parameterised by two slopes $S_\alpha$ and $S_\delta$ along right ascension and declination respectively. 

\begin{eqnarray*}
v|T=\mathbf{o},\alpha,\delta & \sim & \mathcal{N}(V_o+\\
	S_\alpha\,(\alpha-\alpha_o)\cos\delta_0 & + & S_\delta\,(\delta-\delta_o), \sigma_o)\\
v|T=\mathbf{h} & \sim & \mathcal{N} (V_h, \sigma_h)\\
v|T=\mathbf{d} &  \sim  & \mathcal{N} (V_d, \sigma_d)\\
{[}Fe/H{]}|T=\mathbf{o} & \sim & \mathcal{N} ({[Fe/H]}_o, \sigma_{[Fe/H],o})\\
{[}Fe/H{]}|T=\mathbf{h} & \sim & \mathcal{N} ({[Fe/H]}_h, \sigma_{[Fe/H],h})\\
{[}Fe/H{]}|T=\mathbf{d} & \sim & \mathcal{N} ({[Fe/H]}_d, \sigma_{[Fe/H],d})
%\textbf{
\end{eqnarray*}

Here the parameters $V_h$, $V_d$, $V_o$, $\sigma_h$, $\sigma_d$, $\sigma_o$ refer to the mean velocity and velocity dispersions of the halo, disk and Hydrus~1 populations respectively, while 
$[Fe/H]_h$,
$[Fe/H]_d$,
$[Fe/H]_o$,
$\sigma_{[Fe/H],h}$,
$\sigma_{[Fe/H],d}$,
$\sigma_{[Fe/H],o}$ refer to the means and dispersions in metallicity for the halo, disk and Hydrus~1.

The last part of the model describes the distribution over types $T$ for the samples of giants and dwarfs, conditional on angular distance  bin from Hydrus~1's center ($D_{bin}$; we use 5 distance bins). We assume that the sample of giants has zero contamination from the MW disk and that the probability of observing a Hydrus~1 member is a monotonically declining function of distance. This reflects prior knowledge  that the stellar density of Hyi~1 member stars with respect to contamination decreases with radius, without assuming a specific density law.  This approach is preferred due to the complexity of taking into account the fiber allocation efficiency (this is similar to the non-parametric approach adopted in the EM-algorithm by \citet{walker09a}).  
 
\begin{eqnarray*}
T|D_{bin}=k,giant &\sim & {\rm Cat}((f_{o,g,k}, 1-f_{o,g,k}, 0))\\
T|D_{bin}=k,dwarf &\sim & {\rm Cat}((f_{o,w,k}, \\
	&&(1-f_{o,w,k})\,f_{h,d}, (1-f_{o,w,k})\,f_{d,w}))
\end{eqnarray*}

The g and w subscripts in the parameters of the model refer to giant and dwarf samples respectively. For example $f_{o,g,k}$ refers to the probability of observing a Hydrus~1 star in the sample of giants in k-th distance bin and $f_{d,w}$ refers to the probability of observing a disk star vs halo star in a sample of dwarfs (we assume that this probability is not dependent on the distance bin).

The model described above assumes error-free velocities and abundances; however, since all our measurements have non-negligible errors, we introduce these uncertainties into the model by assuming that all observed velocities and metallicities are normally distributed around their true values: $v_{obs} \sim {\mathcal N}(v,\sigma_{v,obs})$ and $[Fe/H]_{obs} {\sim \mathcal N}([Fe/H], \sigma_{[Fe/H],obs})$ where $v_{obs}$, $[Fe/H]_{obs}$ are the observed velocities and abundances, $\sigma_{v,obs}$, $\sigma_{[Fe/H],obs}$ are the observational uncertainties  and $v$ and $[Fe/H]$ are true error-free parameters of each star. 

The priors on the parameters of the model are mostly uninformative, 
$V_h \sim U(-300,300)$, 
$V_{d} \sim U(-300, 300)$, 
$V_o \sim U(60,100)$,
$\sigma_h \sim U(50,300)$, 
$\sigma_d  \sim U(10,300)$,
$\sigma_o \sim U(0,15)$,
$[Fe/H]_h \sim U(-5,0)$,
$[Fe/H]_{d} \sim U(-1,0)$, 
$[Fe/H]_o \sim U(-5,0)$, 
$\sigma_{[Fe/H],h} \sim U(0,2)$,
$\sigma_{[Fe/H],d} \sim U(0,1)$,
$\sigma_{[Fe/H],o} \sim U(0,2)$,
$S_\alpha \sim U(-50,50)$,
$S_\delta \sim U(-50,50)$, and $U(0,1)$ on all object type fractions $f_{d,w}$, $f_{o,g,k}$, 
$f_{o,w,k}$.

\begin{figure}
	\includegraphics[width=\columnwidth]{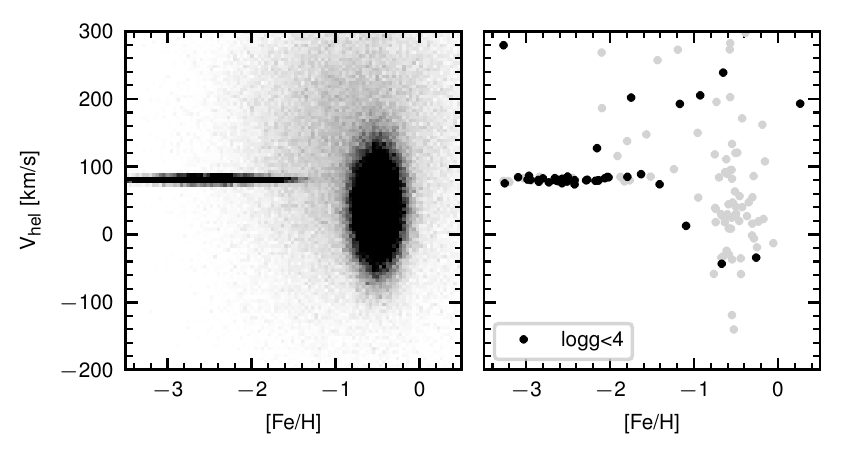}
    \caption{
    The demonstration of our chemo-dynamical model. {\it Left panel:} The 2-D density of stars in metallicity and radial velocity space showing our 3-component Gaussian mixture model. {\it Right panel:} The mock sample of 136 stars from our model, including both the giants $\log g<4$, $\sigma_{\log g}<1$ (black points) and dwarfs (grey points). The figure is meant to be very similar to the fourth panel of Figure~\ref{fig:spec_plot}. The observational uncertainties on both abundances and velocities have not been added to the simulated data points.}
    \label{fig:fake_sample}
\end{figure}

This chemo-dynamical model is implemented in the probabilistic language STAN \citep{carpenter17} and is available in the supplementary materials for this paper.  We run the model on all stars with velocity uncertainties less than 10 km\,s$^{-1}$ (136 in total). We sampled the posterior using a Hamiltonian Monte-Carlo technique \citep{neal12} and No-U-Turn-Sampler \citep{hoffman11}. After running 24 parallel chains for 10000 iterations each, we ensured that the resulting chains are in convergence using the Gelman-Rubin diagnostic \citep{gelman12} for each parameter as well as inspecting the marginal 2-D and 1-D posteriors, all of which show a single, smooth, well-behaved peak. To check that our model provides a good description of the data, we created replicated datasets from our best-fit model. These are shown in Figure~\ref{fig:fake_sample}. The left panel of the Figure shows the overall smooth density in metallicity vs velocity space, as described by our model (assuming the same ratio between giant and dwarf subsamples as in our data), while the right panel shows an exact simulated replica of our dataset with 136 stars. The replication seems to resemble the actual dataset shown in Figure~\ref{fig:spec_plot}, providing some assurance that our model provides an adequate description of the data. 

The numerical results of the chemo-dynamical model fit, i.e. the values of parameters and their uncertainties relevant for the Hydrus~1 system, are  provided in Table~\ref{tab:measurements}, while Table~\ref{tab:chemodyn_extra} lists the parameter values for the contaminant components (disk and halo). Another output from our model is the membership probability for each star. We express these probabilities the form of log-odds of being a member star $log(P_{memb}/P_{nonmber})$ and we include them as a column in Table~\ref{tab:spec_stars}. The uncertainties in those probabilities are obtained after marginalizing over all model parameters. We stress that membership probabilities
for individual stars are often highly uncertain and model dependent; therefore, outputting just a single value of membership probability
without uncertainty as it is common in the literature is meaningful only if model parameters (such as velocity dispersion or systemic velocity) are known precisely (which is almost never true).

\begin{table}
\begin{center}
\begin{tabular}{c|c|c}
\hline
Parameter & Value & Unit \\
\hline
$V_{halo}$ & $142\pm 22$ & km\,s$^{-1}$\\
$\sigma_{V,halo}$ & $134_{-14}^{+17}$ & km\,s$^{-1}$\\
$[Fe/H]_{halo}$ & $-0.83\pm 0.13$ & dex \\
$\sigma_{[Fe/H],{halo}}$ & $0.74\pm 0.1$ & dex\\
$V_{disk}$ & $38.2\pm 7.6$ & km\,s$^{-1}$ \\
$\sigma_{V,disk}$ & $49.5_{-5.7}^{+6.7}$ & km\,s$^{-1}$\\
$[Fe/H]_{disk}$ & $-0.52\pm 0.03$ & dex\\
$\sigma_{[Fe/H],{disk}}$ & $0.17\pm 0.04$ & dex\\
\hline
\end{tabular}
\caption{Measurements of the disk and halo parameters from the chemo-dynamical model}
\label{tab:chemodyn_extra}
\end{center}
\end{table}

Now we briefly summarize the main measurements from the chemo-dynamical model, with more detailed discussion in Section~\ref{sec:discussion}. The systemic heliocentric line-of-sight velocity of Hydrus~1 is $80.4\pm0.6$\, km\,s$^{-1}$, which corresponds to a Galactic Standard of Rest velocity of $-94.1$\, km\,s$^{-1}$.  Given that the angle between the line of sight of Hydrus~1 and the line connecting the Galactic center to Hydrus~1 is only 17 degrees it is likely that the satellite is now on its way toward  perigalacticon.  The other key parameter in the chemo-dynamical model is the velocity dispersion of Hyi~1, which is resolved and measured with high accuracy: $\sigma_v = 2.69_{-0.43}^{+0.51}$\, km\,s$^{-1}$. 
Assuming that the observed stellar velocities cleanly trace Hyi~1's gravitational potential, the total dynamical mass within the projected half-light radius of Hyi~1 can then be estimated using the estimator of \citet{walker09b} $M(<R_{1/2}) \sim 2.2\times 10^5\, M_\odot$. This value corresponds to a mass-to-light ratio of $66_{-20}^{+29}$ solar units. Since  mass-to-light ratio uncertainties are heavily asymmetric, we also compute the logarithm $\log_{10} M/L=1.82 \pm 0.16$, for which the uncertainties are close to Gaussian. As this ratio is clearly higher than that of the stellar population alone, this confirms suspcions based on morphology and structural parameters: Hydrus~1 is a dwarf galaxy dominated by dark matter.

The mean metallicity and metallicity dispersion of Hyi~1 are $[Fe/H]=-2.52\pm 0.09$ and $0.41\pm 0.08$, respectively. The average metallicity is consistent with the metallicity of other ultra-faint dwarfs and agrees with the stellar-mass/metallicity relation traced by other dwarf galaxies \citep{kirby2013}. We also note that the scatter in metallicities within Hydrus~1 is resolved and is significantly larger than zero, which is another indicator of the dwarf galaxy nature of the object \citep[see e.g.][]{willman12}.

\begin{figure}
	\includegraphics[width=\columnwidth]{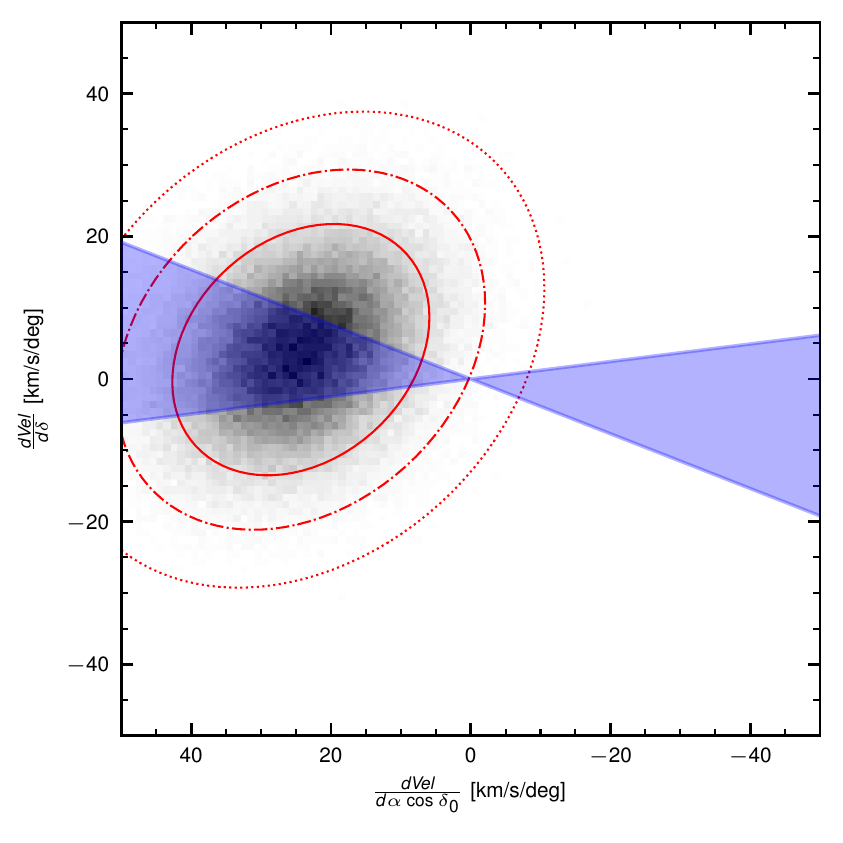}
    \caption{2-D posterior density of the  spatial line-of-sight velocity gradient as inferred from the M2FS kinematic data. The ellipses represent the posterior volume corresponding to 1, 2 and 3 $\sigma$. The blue wedge shows the range of direction of possible velocity gradients if they were aligned along the major axis of Hydrus~1. The width of the wedge reflects the uncertainty on the measurement of the positional angle of Hyi~1's major axis.}
    \label{fig:vel_grad}
\end{figure}

The best fit values for the line-of-sight velocity gradient in Hydrus~1 are $S_\alpha=\frac{dV_{hel}}{d\alpha \cos \delta_{cen}}=24.4\pm 9.6$\,km\,s$^{-1}$\,deg$^{-1}$, and $S_\beta=\frac{dV_{hel}}{d\delta}=4.1\pm 9.2$\,km\,s$^{-1}$\,deg$^{-1}$. We notice that the gradient along right ascension is marginally different from zero (at $\sim 2.5$ sigma significance level). To investigate this further, we show the 2-D posterior on the velocity gradient in Figure~\ref{fig:vel_grad}. We also show there the wedge that would correspond to the radial velocity gradient along the major axis of the system (the width of the wedge represents our uncertainty on the position angle of the major axis). We note that the marginal velocity gradient signal seems to be aligned with the major axis of Hydrus~1. We  discuss the implications of this result in more detail in Section~\ref{sec:discussion_gradient}.

\subsection{Chemical abundances}
\label{sec:chemistry}

As shown in the previous section, our spectroscopic dataset contains around 30 members of Hydrus~1, many of which are quite bright ($r<18$) and have high-quality spectra. Therefore we synthesised spectra of 27 high signal-to-noise (S/N>5) and high probability ($P_{Hyi 1}/P_{bg}>1$) Hydrus~1 stars in order to estimate their detailed chemical abundances. The observed wavelength region includes transitions of Mg (the  Mg~Ib triplet), C (the C$_2$ Swan band), Fe, and to a lesser extent, Ti and Ni. We used the stellar parameters $T_{\rm eff}$, $\log{g}$, and $[{\rm Fe}/{\rm H}]$ estimated from the previous Section, with microturbulence $v_{t}$ estimated using the empirical relationship from \citet{Kirby_2008}. We used the \citet{Castelli_Kurucz} plane-parallel model photospheres, and compiled a line list with atomic transitions from the Gaia-ESO Survey  \citep{Heiter} and molecular transitions (primarily C$_2$) from \citet{Ram_2014}\footnote{Available in \texttt{MOOG} format from \url{http://www.as.utexas.edu/\~chris/lab.html}}. We used the 2014 July version of the \texttt{MOOG} spectral synthesis program \citep{Sneden_1973}.

\begin{figure*}
	\includegraphics[width=\textwidth]{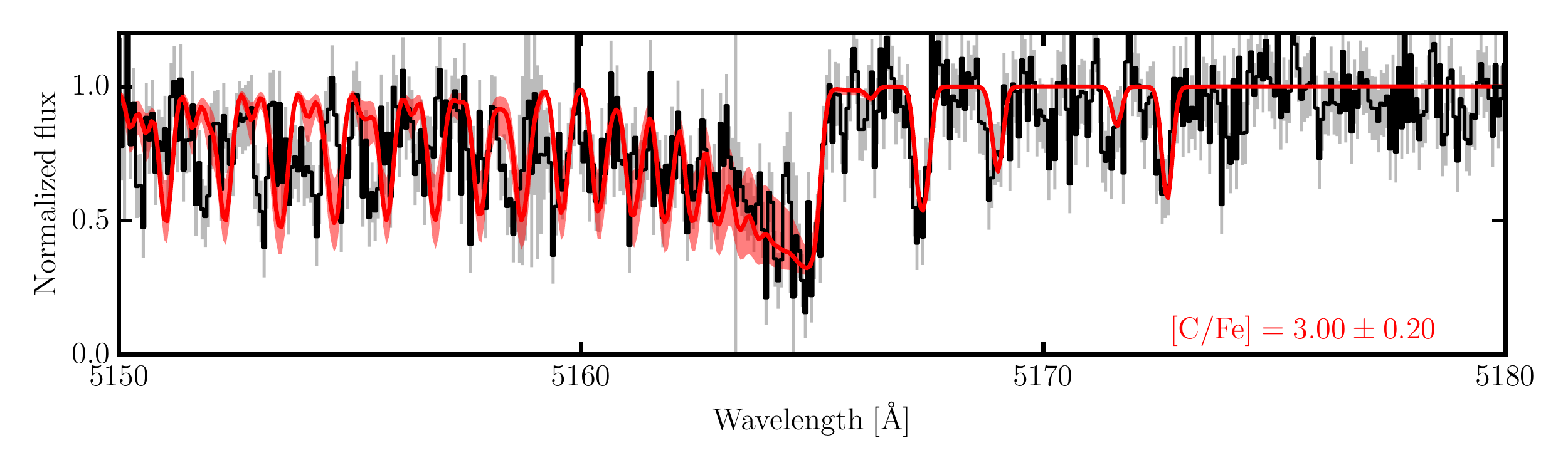}
	\caption{Spectrum of the carbon enhanced extremely metal-poor (CEMP) Hydrus~1 star (star 75) with spectral synthesis model overplotted in red. Gray shading shows the observational uncertainties in the measured spectrum and red shading shows the uncertainties associated with the spectral synthesis parameters. The spectrum shows a strong carbon absorption band and low metallicity indicative of CEMP stars.}  
\label{fig:carbon_star}
\end{figure*}

We masked regions with stellar absorption lines and used a spline function to continuum-normalise the observations \citep{Casey_2014}. Continuum normalisation was relatively straightforward for all but one star, Star 75, which we found to be strongly enhanced in carbon, producing significant absorption at wavelengths bluer than 5165\,\AA. Carbon abundances could only be estimated for Star 75, where we adopted a $^{12}$C/$^{13}$C ratio of $3.5$, and we report upper limits on [C/H] for most other members. Despite using a C$_2$ molecular line list that was compiled specifically for accurate predictions of carbon from the Swan C$_{2}$ bands, we found that there were significant discrepancies between the predicted and observed wavelengths of the C$_2$ feature near 5165\,\AA\ in Star 75. Although the S/N in this spectrum is relatively low ($S/N \approx 6$\,pixel$^{-1}$), the predicted `saw-tooth' spectral features seem to almost grow out of phase with the observations (Figure \ref{fig:carbon_star}), even after allowing for a small residual radial velocity shift to fit the observations. Adopting different isotopic fractions of $^{12}$C/$^{13}$C did not resolve the issue. Discrepancies in wavelength position of this bandhead have been noted in other works \citep{Brooke_2013,Ram_2014}, but not to the extent that seems apparent in Star 75. On the other hand, the low S/N ratio does make it difficult to properly quantify the extent of the wavelength discrepancies. The estimated abundance ratio of $[{\rm C}/{\rm Fe}] = +3$ seems to be largely independent of these wavelength offsets, because if we convolve the observations and data with a Gaussian kernel to smooth out the spectral features and produce a near-continuous opacity blueward of 5165\,\AA\ that is due to the C$_2$ Swan band, the model and observations fit excellently with $[{\rm C}/{\rm Fe}] = +3$. The molecular features, and the level of near-continuous opacity when smoothed, is extremely sensitive to the level of [C/H], so the estimated abundance ratio of [C/H] seems to be largely insensitive to the differences in the wavelengths of the C$_2$ bandhead. 

We were able to estimate [Mg/H] abundances for all Hydrus~1 members. The Ti and Ni atomic lines in this wavelength region are relatively weak, only allowing for one measurement of $[{\rm Ni}/{\rm H}] = -2.15 \pm 0.11$ for Star 154, and upper limits  of $[{\rm Ni}/{\rm H}] < -2.68$ for Star  165, and $[{\rm Ti}/{\rm H}] < -2.46$ for Star 52. Although these limits are not  very informative, we include them here for completeness. We were able to estimate Fe abundances by spectral synthesis for most stars, except for those with very low S/N ratios (e.g., S/N $\approx 5$\,pixel$^{-1}$) or very hot stars ($T_{\rm eff} \approx 8000$\,K). These estimates are in very good agreement with those estimated by template fitting in
Section~\ref{sec:spec_data}: well within the estimated uncertainties of either method. Uncertainties in chemical abundances were estimated during the fitting process by projecting the derivatives in the synthesised spectra with the uncertainties in the continuum-normalised fluxes. The estimated abundances, uncertainties, and upper limits, are given in Table \ref{tab:chemical_abundances}. We discuss these measurements in the next section of the paper.

\begin{table}
\begin{center}
\begin{tabular}{|l|c|c|c|c|c|}
\hline
Star ID & [C/H]             &  [Mg/H]               & [Fe/H]              \\ 
\hline
  4     & $<-0.32 $         &  $-2.64 \pm 0.11$     & $-2.42 \pm 0.10$    \\
 26     & $<-0.15 $         &  $-3.04 \pm 0.15$     & $-2.15 \pm 0.10$    \\
 30     & $<-0.74 $         &  $-2.27 \pm 0.10$     & $-1.74 \pm 0.10$    \\
 34     & $<-0.64 $         &  $-2.72 \pm 0.12$     & $-2.94 \pm 0.10$    \\
 49     & $<-0.82 $         &  $-2.38 \pm 0.10$     & $-2.97 \pm 0.10$    \\
 52     & $<-1.13 $         &  $-1.78 \pm 0.10$     & $-2.38 \pm 0.10$    \\
 61     & $<-0.43 $         &  $-2.07 \pm 0.16$     & $-3.08$ \\
 68     & $<-1.03 $         &  $-2.40 \pm 0.10$     & $-2.67 \pm 0.10$    \\
 69     & $<-0.71 $         &  $-2.26 \pm 0.11$     & $-2.61 \pm 0.10$    \\
 72     & $<-0.74 $         &  $-2.53 \pm 0.11$     & $-2.80 \pm 0.14$    \\
 75     & $-0.18 \pm 0.20$  &  $-2.77 \pm 0.14$     & $-3.18$ \\
 77     & $<-1.06 $         &  $-1.01 \pm 0.10$     & $-1.60 \pm 0.10$    \\
102     & $< 0.98 $         &  $-2.61 \pm 0.13$     & $-3.17 \pm 0.18$    \\
119     & $<-0.47 $         &  $-2.43 \pm 0.12$     & $-2.87 \pm 0.10$    \\
139     & $       $         &  $-2.13 \pm 0.10$     & $-2.26$ \\
149     & $< 1.82 $         &  $-0.70 \pm 0.12$     & $-1.40$ \\
150     & $<-1.28 $         &  $-2.33 \pm 0.10$     & $-2.52 \pm 0.10$    \\
151     & $       $         &  $-2.28 \pm 0.34$     & $-2.31$ \\
154     & $<-1.43 $         &  $-2.34 \pm 0.10$     & $-2.55 \pm 0.10$    \\
165     & $<-1.58 $         &  $-2.55 \pm 0.10$     & $-2.85 \pm 0.10$    \\
182     & $<-1.01 $         &  $-2.57 \pm 0.10$     & $-3.01 \pm 0.10$    \\
188     & $       $         &  $-2.26 \pm 0.29$     & $-2.23 \pm 0.11$    \\
209     & $<-0.12 $         &  $-2.36 \pm 0.14$     & $-3.12 \pm 0.13$    \\
221     & $       $         &  $-1.84 \pm 0.10$     & $-2.21$ \\
222     & $< 0.07 $         &  $-2.24 \pm 0.23$     & $-2.68 \pm 0.12$    \\
225     & $<-0.71 $         &  $-2.59 \pm 0.10$     & $-2.54 \pm 0.10$    \\         
\hline
\end{tabular}
\end{center}
\caption{Chemical abundances and upper limits estimated via spectral synthesis (Section \ref{sec:chemistry}) for high probability members of Hyi~1.}
\footnotetext{}
\label{tab:chemical_abundances}
\end{table}

\section{Discussion}
\label{sec:discussion}

In this section we discuss and summarize the key results from the photometric, chemical and dynamical analyses of Hydrus~1 presented in Sections~\ref{sec:sat_properties} and \ref{sec:spectroscopy}. In particular we focus on possible radial velocity gradient of the dwarf, its dark matter content, chemical properties, connection to the Magellanic Clouds and dark matter annihilation searches.

\subsection{Radial velocity gradient}
\label{sec:discussion_gradient}
\begin{figure}
	\includegraphics[width=\columnwidth]{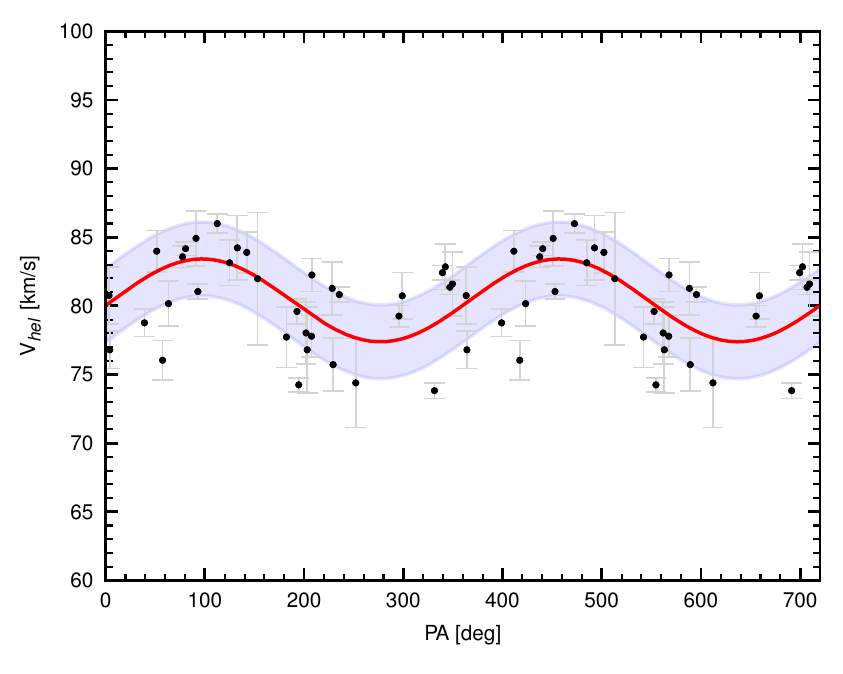}
    \caption{Line-of-sight velocities of Hydrus~1 stars  vs position angle. Only stars with velocity uncertainty less than 5\,km\,s$^{-1}$ and [Fe/H]<-1 are shown. All the data points have been plotted twice, once with the original position angles, and once with position angles offset by 360 degrees. The red curve shows the radial velocity change according to the best fit velocity gradient evaluated at half-light radius (corresponding to $\sim$ 3\,km\,s$^{-1}$ rotation velocity). The shaded region around the curve has the half-width of the best fit velocity dispersion of the system.}
    \label{fig:rotation}
\end{figure}

The radial velocity gradient that we detect in Hyi~1 is only  marginally significant, at 2.5 $\sigma$, and seems to be aligned with the major axis of the system.  If we assume that it is not a statistical fluctuation, possible  causes of this gradient would include satellite rotation, perspective rotation and tidal disruption. 
Perspective rotation \citep{kaplinghat08,walker08} is caused by projection effects from tangential motion of an object with large angular size. However as Hyi~1's size is only $\sim 0.1\degr$\ on the sky, the observed gradient of $\sim
24$\,km\,s$^{-1}$\,deg$^{-1}$  would imply an unrealistically high tangential velocity ($\gtrsim 1000$\,km\,s$^{-1}$), and is therefore an unlikely cause of the observed gradient. Another possible explanation is tidal disruption of the satellite. This has been claimed to be observed in a few MW satellites 
\citep{aden09,collins17}, albeit at low levels of significance. We note however that the observed gradient is much larger than expected from a simple velocity gradient along the orbit in the MW gravitational potential. Therefore, in order to explain the gradient in terms of disruption, significant fine-tuning of the orbit would be required \citep{kupper10,martin10,kupper17}. 
Furthermore, the low ellipticity of Hyi~1 and the apparent regularity of the satellite and its density profile argue against this possibility. Finally, the velocity gradient in the system could be explained by satellite rotation. 
Assuming the measured radial velocity gradient along the major axis corresponds to solid-body rotation,  we estimate a rotation velocity of $3$\,km\,s$^{-1}$ at the half-light radius, which is comparable to the velocity dispersion of the system. 
To illustrate the plausibility of this interpretation, we show in Figure~\ref{fig:rotation} the radial velocity of possible Hydrus~1 members (selected based on metallicities only) as a function of position angle of the system, with the curve showing the line-of-sight velocity model evaluated at the half-light radius (essentially interpreted as satellite rotation). 
We note that the model seems to fit the data well, matching the observed RV changes as a function of positional angle. If the rotation hypothesis is a correct interpretation of the data, that would make Hydrus~1 the first ultra-faint dwarf with significant rotation. 
Why would be Hydrus~1 be an exception?  It is difficult to say for sure, but it is possible that at least some of the dwarf satellites of the MW had  disk-like morphology before accretion, and were then transformed into spheroidal galaxies by a process  known as tidal stirring \citep{mayer01,kazantzidis11}.  In this case some of the satellites could still be in the transformation phase with significant rotational support. That would mean that Hydrus~1 became a satellite very recently.

However, due to the low statistical significance of the rotation signal, we urge caution and emphasise the need for more data to confirm or refute it.

\subsection{Nature of Hydrus 1}

While the nature of Hydrus~1 as a dwarf galaxy is established by its size, dark matter content and internal metallicity distribution, in this Section we put Hyi~1 in context with other systems of similar structural and chemo-dynamical properties.  

\begin{figure}
	\includegraphics[width=\columnwidth]{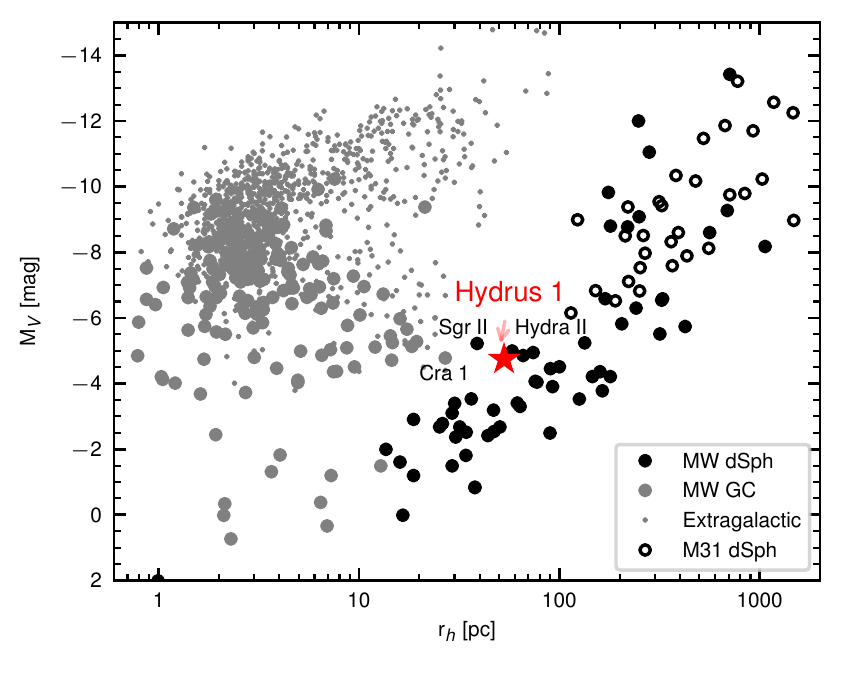}
    \caption{Distribution of sizes and luminosities of various faint stellar systems. Hydrus~1 is shown by a red star. MW/M31 dwarf spheroidal galaxies are shown by black filled/empty circles,  respectively. Large grey circles show MW globular clusters, while small grey circles are showing extra-galactic compact stellar systems ($<100$\,pc) from \citet{brodie11}. Data are taken from Figure 6 of \citet{torrealba18} and include the most recent dwarf galaxy and globular cluster discoveries.}
    \label{fig:lum_size}
\end{figure}

\begin{figure}
	\includegraphics[width=\columnwidth]{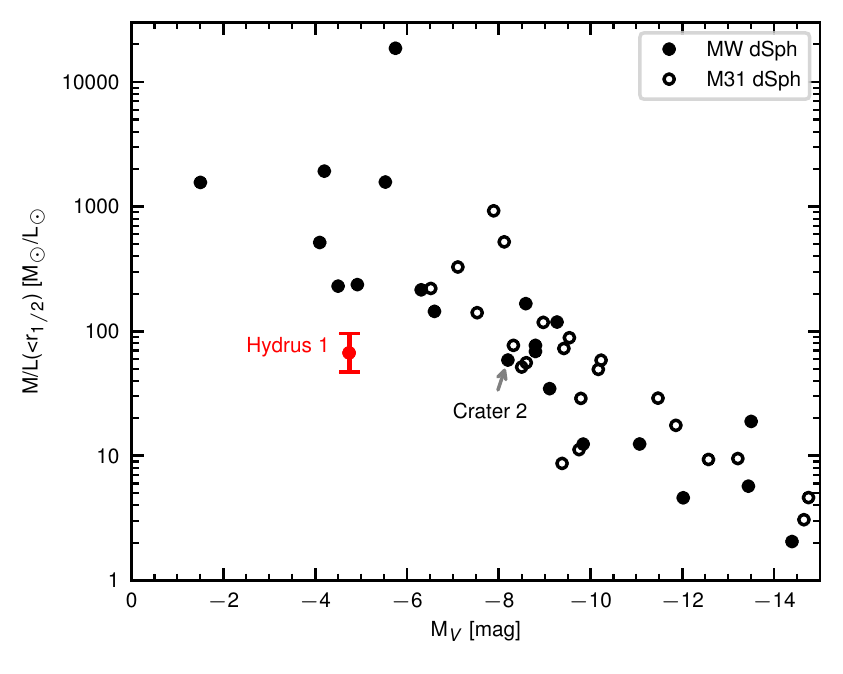}
    \caption{Dynamical mass-to-light ratio as a function of luminosity for dwarf satellites of the MW and M31. Hyi~1 is marked by a red circle. Data come from the 2015 version of the dwarf galaxy database by \citet{mccon12} as well as additional data by \citet{caldwell17,li18}. }
    \label{fig:mass_to_light}
\end{figure}

With a circularised effective radius of $\sim$ 50\,pc and luminosity of $M_V\sim-4.7$, Hydrus~1 sits in an ambiguous between dwarf galaxies and extended globular clusters (GCs) in the luminosity-size space. Figure~\ref{fig:lum_size} shows all known MW satellites (dSphs and GCs) together with Hydrus~1. The objects similar in size and luminosity to Hydrus~1 are the Hydra~II satellite \citep{martin15} with a luminosity of $M_V=-4.8$ and size $r_h=68$\,pc, and the Sagittarius~II satellite \citep{laevens15} with a luminosity of $M_V=-5.2$ and half-light radius of $\sim 37$\,pc. The Crater~1 (or Laevens~1)  satellite \citep{belokurov14,laevens14} has similar luminosity ($M_V=-5.3$) to Hydrus~1, but significantly smaller size ($\sim$ 20\,pc). 
The classification of Hydra~II is somewhat unclear because its velocity dispersion is only marginally resolved; however, \citep{kirby15} argue that its chemical abundances favor a dwarf galaxy interpretation. Sagittarius~II does not have any velocity dispersion measurement. Finally, the Crater~1 satellite currently seems to be classified as an extended globular cluster based on the star-formation history and unresolved velocity dispersion \citep{weisz16,voggel16}. We note however that the mass-to-light ratio of Crater~1 from \citet{voggel16} $M/L=8.52_{-6.5}^{+28.0}$ is formally consistent with both the absence of dark matter and a significant amount of dark matter. 

While the mass-to-light ratio measurement of $\log M/L = 1.82\pm 0.16$ in Hyi~1 unambiguously indicates the presence of dark matter, it is worth comparing Hyi~1 to other dwarfs of similar luminosity. Figure~\ref{fig:mass_to_light} shows mass-to-light ratio as a function of luminosity for MW and M31 satellite dwarf galaxies; Hydrus~1 is identified by a red marker. We note that Hyi~1 seems to be an outlier, possessing a mass-to-light ratio well below (by a factor of 3) other galaxies  with similar luminosities. What makes Hydrus~1 such an outlier? 
 One possibility is that some of the faint objects like Hydra~II, Crater~1 and Hydrus~1 and others that occupy the area in size-luminosity space between globular clusters and dwarf galaxies could be intermediate-type objects with small but nonzero dark matter content and mass-to-light ratios smaller than other typical ultra-faint dwarfs. The origin of these systems could then be related either to tidal disruption after accretion \citep{penarrubia08} or to their initial formation \citep{ricotti16}.  Testing this hypothesis would be quite difficult, however, as the known objects near the dwarf galaxy/globular cluster interface have small velocity dispersions and we lack high enough quality data on these systems to rule out a smooth transition from dark matter dominated objects to dark matter-free globular clusters.  

Another possible explanation for the apparent low mass-to-light ratio of Hyi~1 could be related to our tentative measurement of the line-of-sight velocity gradient (see Section~\ref{sec:spec_modeling} and \ref{sec:discussion_gradient}). If the gradient is real and is caused by galaxy rotation, the mass estimator based on velocity dispersion alone could be underestimating the dynamical mass and therefore the mass to light ratio. To gauge the possible impact of this we make a simple modification of the \citet{walker09b} estimator, adding a rotation term
$$ M(<r_{1/2})  = \frac{5}{2 G} \left(\sigma^2 + \frac{2}{5} V_{rot}^2\right) r_{1/2} $$ 
Assuming  that the rotation velocity can be obtained from the radial velocity gradient  evaluated at half light radius 
$V_{rot} = \frac{dV}{d\alpha \cos \delta_{cen}} r_{1/2}$ and that we observe the rotation in the edge-on orientation, we can obtain a half-light dynamical mass of 
$ M(<r_{1/2}) = 2.60_{-0.70}^{+0.98}\, 10^{5}$\, M$_\odot$. This measurement is larger by $\sim$ 20\% than the dynamical mass calculated without taking into account the rotation and would imply the mass-to-light ratio of $\sim 80$. This is not enough, however, to put Hydrus~1 in line with other ultra-faints on Fig.~\ref{fig:mass_to_light}. Of course, the impact of rotation can be increased if we assume that the rotation plane is not viewed edge-on, in which case the true rotation would be higher by $\cos i$ and could in principle increase the dynamical mass of Hydrus~1 to match the overall relation in Fig.~\ref{fig:mass_to_light}.

\subsection{Connection to the Magellanic Clouds}

\begin{figure}
	\includegraphics[width=0.48\textwidth]{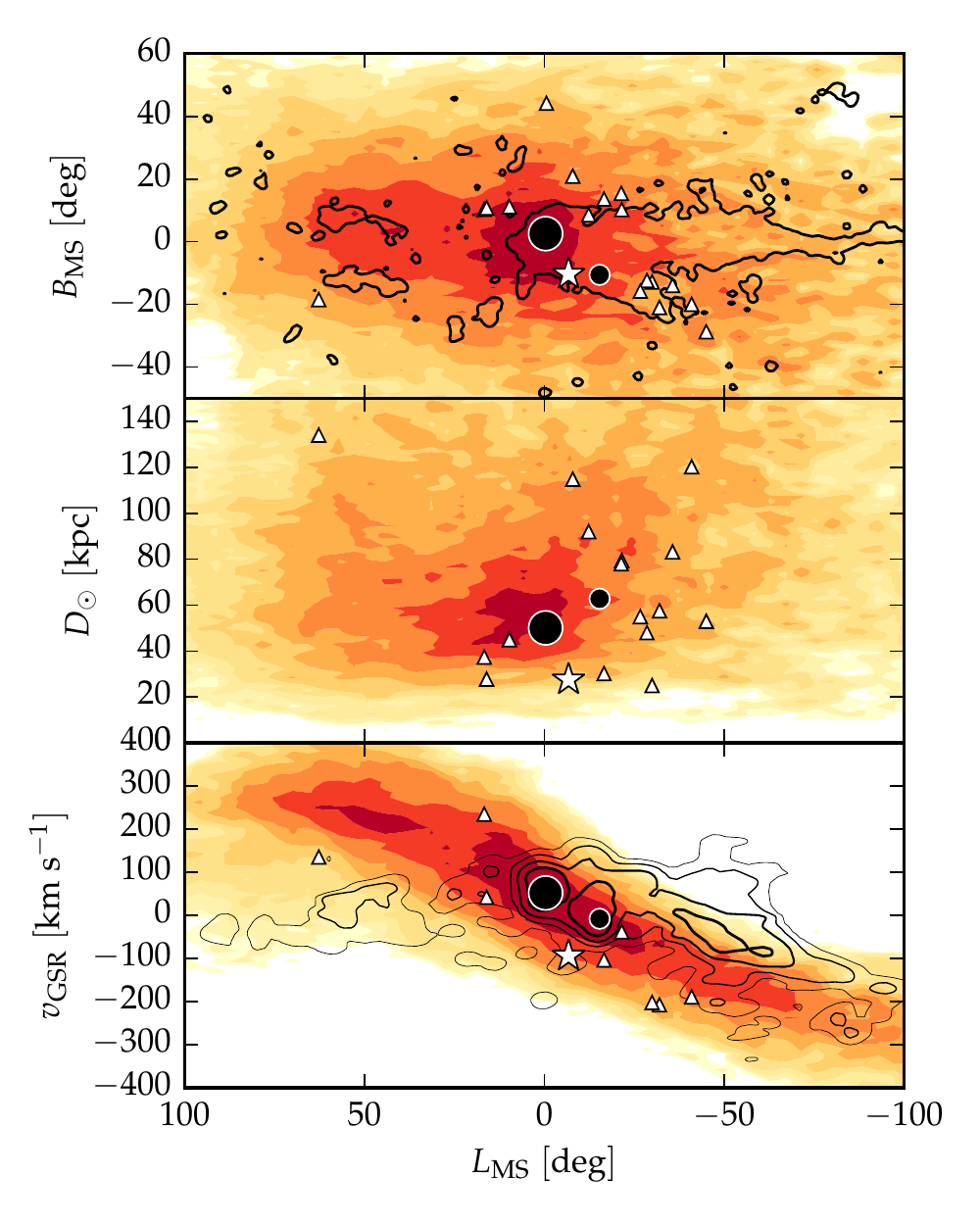}
    \caption{Density of the LMC debris in the maximum-likelihood model
      of \citet{jethwa16}. {\it Top:} Probability distribution of the
      LMC satellites on the sky in the Magellanic coordinate system as
      defined by \citet{nidever08}. Black contours show neutral
      hydrogen gas in the Magellanic Stream. Large and small black
      circles mark the locations of the LMC and the SMC
      respectively. White star marker shows the location of Hydrus~1,
      while white triangles show the location of known MW
      satellites. {\it Middle:} Colored density contours represent
      the expected probability distribution of the LMC satellites in
      the Magellanic latitude vs heliocentric distance space. The
      meaning of the symbols is the same as in the top panel. The
      contours for the Magellanic stream are not shown due to the lack
      of distance estimates available. {\it Bottom:} Same as above but
      for the galactocentric line-of-sight velocity vs Magellanic
      latitude.}
    \label{fig:mc_disruption}
\end{figure}

As seen in Figure~\ref{fig:lmc_map}, the location of the Hydrus~1
satellite on the sky is right in between the Large and the Small
Magellanic Clouds. Moreover, as discussed in
Section~\ref{sec:spatial_distrib}, the satellite orientation on the
sky appears to be aligned with the direction of the Clouds' motion on
the sky. Thus, it is not imprudent to ask: how likely is the
association of Hydrus~1 with the Magellanic accretion event? In fact,
the Magellanic hypothesis has recently been invoked to explain the
nature of many of the satellites discovered in the vicinity of the
Clouds in the past few years
\citep{koposov15,koposov15b,drlica_wagner15,torrealba18}. Unfortunately,
it is impossible to give the exact answer to this question without the
complete set of the satellite's phase-space coordinates combined with
the accurate mass estimates of the LMC and the SMC. Nonetheless, one
may draw some useful conclusions by comparing the numerical
simulations of the Magellanic Clouds' disruption to the observed properties
of the satellite \citep[see e.g.][]{nichols11,jethwa16,sales17}.

\begin{figure*}
  \includegraphics[width=\textwidth]{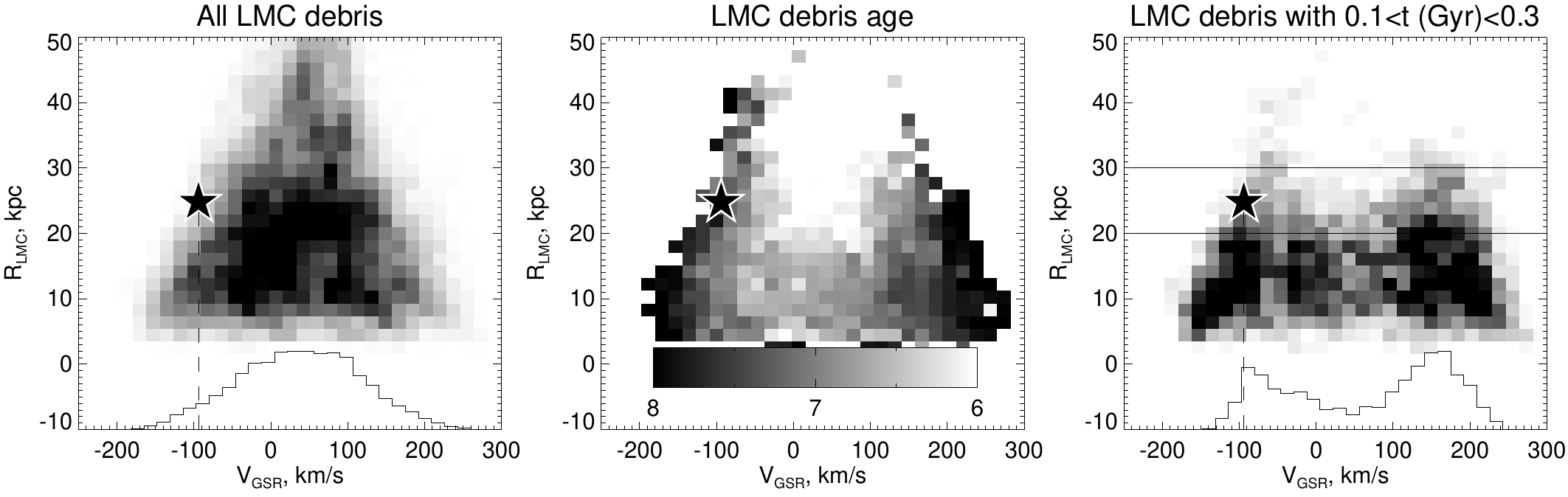}
  \caption{Distribution of the line-of-sight velocity $V_{\rm GSR}$ of
    the simulated LMC debris as a function of the distance from the
    Cloud, $R_{\rm LMC}$. In this particular simulation, the
    MW/LMC/SMC masses are fixed at 75/5/4$\times10^{10} M_{\odot}$. We
    select particles at the same Magellanic longitude $L_{\rm MS} \sim
    -7^{\circ}$ as that of Hyi 1, but in a range of latitudes $B_{\rm
      MS}$. The location of the Hyi 1 satellite is marked with a black
    star. {\it Left:} Density of particles in the $V_{\rm GSR}, R_{\rm
      LMC}$. The black histogram shows the 1D distribution of $V_{\rm
      GSR}$. {\it Middle:} Mean dynamical age of the particles as a
    function of $V_{\rm GSR}$ and $R_{\rm LMC}$. According to the
    greyscale bar displaying the logarithm of the stripping time in
    years, the bulk of the debris was removed from the LMC less than
    0.5 Gyr ago. There is a clear age gradient as a function of
    $V_{\rm GSR}$, with older debris in the leading (trailing) tail
    possessing higher negative (positive) velocities. {\it Right:}
    Only debris stripped between 0.1 and 0.3 Gyr ago are
    shown. Histogram shows the 1D $V_{\rm GSR}$ distribution for
    particles with $20<R_{\rm LMC}/1$\,kpc$<30$, i.e. those at
    distances from the LMC similar to that of Hyi 1. Note the strong
    bimodality of the velocity distribution.}
  \label{fig:mc_age}
\end{figure*}

Figure~\ref{fig:mc_disruption} shows the expected distributions of the
particles stripped from the Large Magellanic Cloud as computed by
\citet{jethwa16}. The top panel gives the positions on the sky in the
Magellanic Stream coordinate system \citep[see][]{nidever08}. The
middle panel presents the dependence of the heliocentric distance on
the Magellanic stream longitude $L_{\rm MS}$. Finally, the bottom row
reveals how the galactocentric radial velocity evolves with $L_{\rm
  MS}$. Over-plotted on top of the distribution of the simulated
particles are the positions of the currently known MW satellites
together with that of Hydrus~1. Unsurprisingly, as illustrated in the
top panels, on the sky, Hyi~1 sits right at the peak of the predicted
debris distribution. However, the match between the simulations and
the observations appears to be much less convincing if the other two
phase-space projections are considered. In particular, Hyi~1 is
significantly closer than the majority of the LMC debris, yet its
heliocentric distance is in marginal agreement with the maximum
likelihood model prediction. Similarly, the line-of-sight velocity
measurement of Hyi~1 is not well reproduced by the model, but appears
compatible with the broad trend of increasing $V_{\rm GSR}$ as a
function of $L_{\rm MS}$. It is difficult to speculate how informative
the agreement between the observed position of Hydrus~1 and the
\citet{jethwa16} models is, given that the width of the LOS velocity
distribution is $\sim$ 200 km\,s$^{-1}$. Naively, this would imply
that there is a $\sim$ 50\% chance that a random MW halo object, not
related to the MC in-fall, will possess a velocity in agreement with
the \citet{jethwa16} predictions.

Some of the thickness of the density contours of the LMC debris in
Figure~\ref{fig:mc_disruption} may be due to the maximum likelihood
model of \citet{jethwa16} attempting to explain the wide spread of the
positions of the DES satellites on the sky and in the distances along
the line of sight. While this model generates a broad cloud of tidal
debris which envelopes most of the satellites, it appears to be
incorrect in detail. As pointed out by \citet{jethwa16}, the satellite
distribution around the LMC is rather anisotropic, with many of the
satellites forming a narrow sequence aligned with the projection of
the SMC's orbital plane onto the sky (see Figure 7 of
\citet{torrealba18}). Therefore, given that the above maximum
likelihood model may be over-bloated, it is useful to conduct a
different experiment and explore instead whether there exists any
model of the Clouds' accretion in which the observed phase-space
position of Hyi 1 is significantly more likely. Motivated by the
recently reported measurements giving preference to the ``light''
Galaxy \citep[see e.g.][]{gibbons14,williams17,eadie17,patel18}, we
focus on a particular simulation, in which the MW/LMC/SMC masses are
75/5/4$\times 10^{10} M_{\odot}$. For this simulation,
Figure~\ref{fig:mc_age} shows the density of the LMC particles in the
plane spanned by the line-of-sight velocity $V_{\rm GSR}$ and the
distance from the LMC, $R_{\rm LMC}$, marginalized over 150 different
values of the Clouds' proper motions. Note that we select particles in
a narrow slice of Magellanic longitudes corresponding to that of Hyi~1
and a range of Magellanic latitudes. The left panel of the Figure
shows the view of the LMC debris not too dissimilar to the ML model
predictions displayed in Figure~\ref{fig:mc_disruption}. Here, a broad
- albeit lumpy - $V_{\rm GSR}$ distribution is visible, without any
strong trends as a function of $R_{\rm LMC}$. However, as the middle
panel of Figure~\ref{fig:mc_age} reveals, there exists a noticeable
dynamical age gradient across the $(V_{\rm GSR}, R_{\rm LMC})$
plane. More precisely, the particles stripped recently, i.e. less than
$0.1$\,Gyr ago populate low $V_{\rm GSR}$ values across a wide range
of the LMC distance. However, the debris unbound from the Cloud more
than $0.1$\,Gyr ago form tight sequences with high negative and high
positive velocities, likely corresponding to the leading and the
trailing tails. Hyi~1 (shown as a black star) can be found on the
leading sequence, whose $V_{\rm GSR}$ velocity decreases away from the
LMC. As displayed in the right panel of the Figure, if only
dynamically old particles are selected, the histogram of their radial
velocities becomes strongly bimodal with Hyi~1 situated near the peak
of the leading debris distribution.

\begin{figure}
  \includegraphics[width=0.48\textwidth]{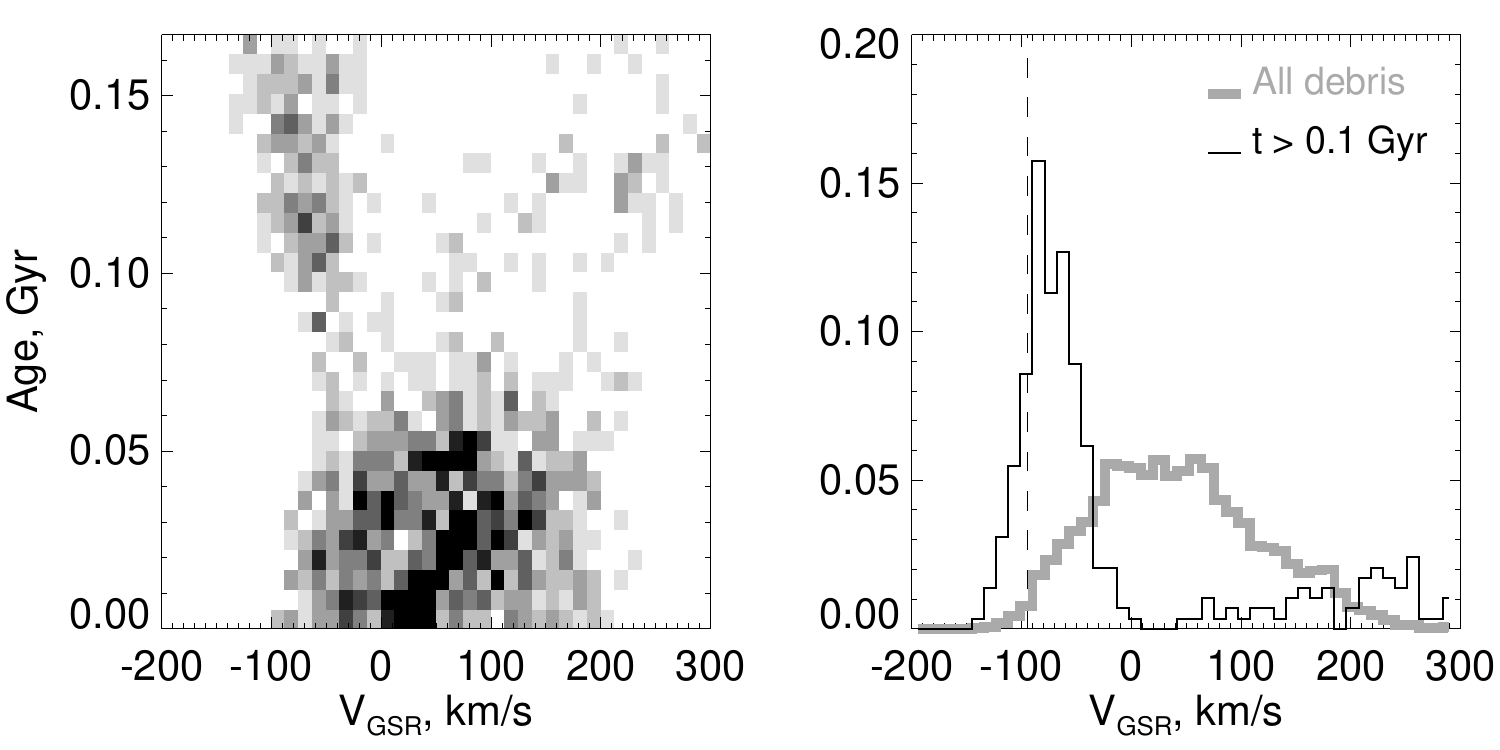}
  \caption{Distribution of the LMC debris in the simulation considered
    in Figure~\ref{fig:mc_age} within 10 kpc of Hyi~1. {\it Left:}
    Density of particles in the plane of Galactocentric line-of-sight
    velocity and dynamical age (time of stripping from the
    LMC). Notice two distinct clouds of debris: the one corresponding
    to the most recently unbound particles, those with age $<0.1$\,
    Gyr and a range of velocities, and the second group with $t>0.1$\,
    Gyr. {\it Right:} 1D velocity distribution of all debris (debris
    with $t>0.1$\,Gyr) shown in grey (black). Dashed line marks the
    measured $V_{\rm GSR}$ of Hyi~1.}
  \label{fig:mc_age_near}
\end{figure}

We confirm that the old debris, i.e. those stripped longer than 0.1
Gyr ago have radial velocities similar to that measured for Hyi~1 by
selecting particles within 10 kpc of the satellite (in 3D). The
distribution of the simulated debirs is shown in
Figure~\ref{fig:mc_age_near}, where two clumps with distinct age
properties are apparent (left panel). The dynamically old debris
occupy very narrow velocity range very close to $V_{\rm GSR}$ of Hyi~1
(right panel of the Figure). It therefore appears that in the low-mass
MW, some of the satellites stripped in the early phases of the LMC
disruption can populate the leading debris tail (with kinematics
similar to that measured for Hydrus) while staying in a close
proximity to the Cloud. In 3D, Hydrus~1 is only $24$\,kpc away from
the LMC. Interestingly, although this is nominally larger than the
recent low bounds on the tidal radius of the Cloud
\citep{mackey16,besla16}, it is nonetheless quite close to it.

The final answer as to the possibility of the association of Hyi~1 and
the LMC will likely be procured using the data from the second data
release of the Gaia satellite, which will provide plenty of the
satellite's members with measured proper motion with individual
uncertainties up to $\sim 0.1$ mas\,yr$^{-1}$. According to the ML
model prediction of \citet{jethwa16}, a satellite stripped from the
LMC and currently observed at the position of Hyi~1 with $V_{\rm
  GSR}\sim -95 \pm 30$ kms$^{-1}$ would have $\mu_{\rm W}=-2.85 \pm
1.7$ mas yr$^{-1}$ and $\mu_{\rm N}=-3.0 \pm 1.8$ mas yr$^{-1}$ (in
the Heliocentric frame). On the other hand, if we select only the
particles contributing to the overdensity of old debris around the
location of Hyi~1 as shown in Figure~\ref{fig:mc_age_near}, we obtain
$\mu_{\rm W}=2.0 \pm 0.6$ mas yr$^{-1}$ and $\mu_{\rm N}=-1.0 \pm 0.8$
mas yr$^{-1}$.

\subsection{Dark matter annihilation}

Hydrus~1's close proximity makes it an interesting target for indirect searches for photons that might be released as products of dark matter particle interactions---e.g., annihilation or decay. In recent years some of the deepest searches and strongest constraints on annihilating dark matter have come from gamma-ray observations of the Milky Way's dwarf spheroidal galaxies, and in particular from the ultrafaint population revealed by deep sky surveys~\citep[e.g.][]{2015PhRvD..91f3515H,2015PhRvD..91h3535G,2015PhRvL.115w1301A,2015PhRvL.115h1101G,2015ApJ...809L...4D,2016PhRvD..93d3518L,2017ApJ...834..110A}. Here we quantify the expected signal from Hydrus~1 and perform a search for gamma-ray emission with data from the Fermi Gamma-ray Space Telescope.

For a given particle physics model (i.e., particle mass, interaction cross section), the flux of the expected photon signal depends on the distribution of dark matter within the source. For annihilation the flux is proportional to the $J$-profile:

$$ \frac{d J(\hat n)}{d \Omega} = \int \limits_{l=0}^{\infty}[\rho(l \hat  n)]^2\, dl,$$ 
where $l$ is the line of sight distance in direction $\hat n$, $\rho(l\hat{n})$ is the dark matter density at location $l\hat{n}$ and $\Omega$ is solid angle.  Following the procedure described in detail by \citet{geringer15}, we use the spherical Jeans equation to estimate the dark matter density profile, $\rho(r)$, of Hydrus~1.  We adopt the sample of 31 stars for which our measurements of position, velocity and metallicity imply membership probability $>0.95$.  Assuming hydrostatic equilibrium, spherical symmetry, constant velocity anisotropy and negligible contamination by unresolved binary stars, we adopt the same five-parameter dark matter halo model and priors as described by \citet{geringer15}.  We also follow the same procedure to obtain the marginalized posterior probability distribution for $\log_{10}[J(\theta)/ (\mathrm{GeV}^2\mathrm{cm}^{-5})]$, where $J(\theta)$ is the integral of $dJ/d\Omega$ over a solid angle corresponding to angular radius $\theta$. In particular, since the kinematic data do not constrain the dark matter halo beyond the orbits of the star in the dataset, we truncate $\rho(r)$ at the radius of the outermost member star (see \citet[Sec.~6.4]{geringer15}). For Hydrus~1 the estimated (deprojected) distance to the outermost member star is 140~pc ($2.6\, r_{half}$), corresponding to a maximum angle $\theta_{\rm max} =0.29\degr$, beyond which $dJ/d\Omega$ is zero. This truncation gives a conservative estimate of the amplitude predicted for an annihilation signal under a given particle physics model.

\begin{figure}
	\includegraphics[width=\columnwidth]{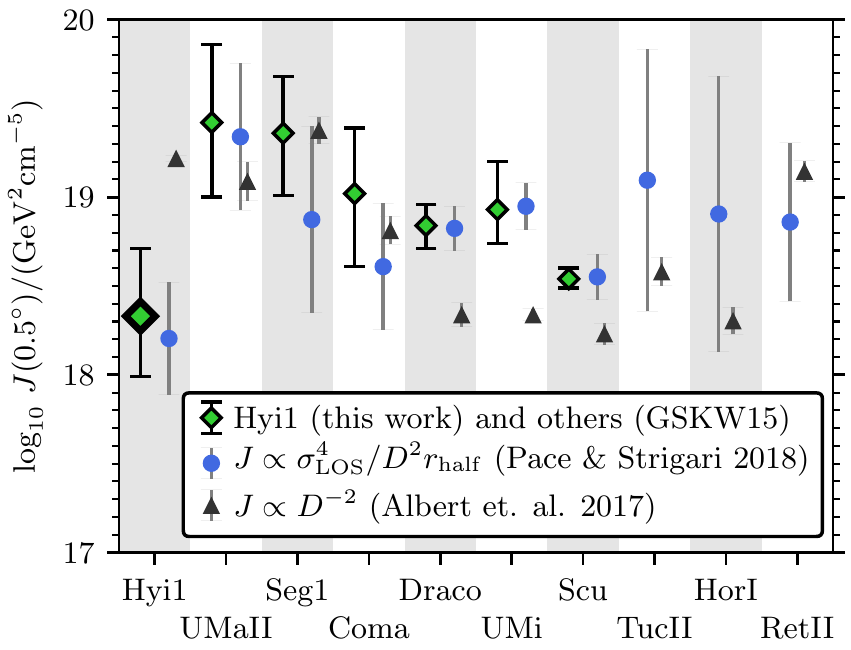}
    \caption{The astrophysical $J$ factor for dark matter annihilation of Hydrus~1 compared with those of other Milky Way dwarf spheroidals. Green diamonds show $J$ of several dwarfs determined with the procedure given in \citet{geringer15} (GSKW15). Blue circles and gray triangles are estimates of $J$ based on simple scaling formulas. They are fit to the GSKW15 $J$'s but their values for Hydrus~1 can be considered predictions.}
    \label{fig:Jcompare}
\end{figure}

The resulting measurement of Hydrus~1's $J$ value is $\log_{10} J(\theta_{\rm max})/(\mathrm{GeV}^2\mathrm{cm}^{-5}) = 18.33^{+0.38}_{-0.34}$ (16th, 50th, and 84th percentiles of the posterior). In Fig.~\ref{fig:Jcompare} we compare Hydrus~1 with Milky Way dwarfs in the ``top tier'' of $J$ values. For purposes of ranking it is important to treat the dwarfs as similarly as possible when computing $J$'s. Different analyses of the same kinematic data set often yield $J$ values differing by amounts comparable to the statistical error. This is due to choices in the Jeans analysis (e.g. form of the dark matter profile, prior ranges, halo truncation, the assumption of spherical symmetry, etc.). In Fig.~\ref{fig:Jcompare} all $J$ values represented by green diamonds are computed using the method~\citep{geringer15}. We omit recently published $J$ values (e.g. those of \citet{2015MNRAS.453..849B,2015ApJ...808L..36B,2015ApJ...808...95S,2016MNRAS.463.1117Z,2016MNRAS.461.2914H,2016MNRAS.463.3630G,walker16,2016MNRAS.462..223B,2017PhRvD..95l3012K,2018arXiv180206811P}) since the comparison with Hydrus~1 will not be straightforward.

There have been several recent attempts to simplify the estimation of $J$ values using functions of photometrically-determined properties (distance, half-light radius) and, sometimes, the total line-of-sight velocity dispersion. These formulas usually have free parameters that are fit to existing compilations of $J$ values based on the Jeans equation. Hydrus~1 presents the first chance to test them. In Fig.~\ref{fig:Jcompare} we show two predictions for the $J$ value for Hydrus~1. Blue circles are based on the formula $J \propto \sigma_{los}^4 D^{-2} r_{half}^{-1}$~\citep{2018arXiv180206811P}, whose form can be motivated by simple analytic considerations~\citep[e.g.][]{2016PhRvD..93j3512E}. In addition to calibrating to their own Jeans analysis of the dwarfs, \citet{2018arXiv180206811P} recalibrate the formula to the \citet{geringer15} data set (green diamonds), finding that it fits dwarf $J$'s with residual error of 0.08 dex. Adopting the latter calibration, we see that the scaling relation appears to fit the measured value for Hydrus~1 fairly well (error bars come from adding in quadrature the uncertainty in velocity dispersion, distance, half-light radius and a 0.08 dex intrinsic scatter). The gray triangles in Fig.~\ref{fig:Jcompare} correspond to a purely photometric determination of $J$, developed by \citet{2015ApJ...809L...4D,2017ApJ...834..110A}, where $J$ is a function of distance alone, scaling as $J \propto D^{-2}$. As with the blue circles, this relation was calibrated to the results of \citet{geringer15}. The distance-based $J$ estimate severely overpredicts Hydrus~1's J value. While the failure of such a simple estimator is perhaps not surprising, the lesson is that spectroscopic observations of dwarf galaxies are necessary to determine their dark matter content. If the distance-based scaling is used to predict dwarf $J$ values it must be accompanied by a large error bar~\citep{2017ApJ...834..110A}. Here, an intrinsic uncertainty $> 0.8$~dex would be needed to bring the estimate  to $< 1\sigma$ tension with the observationally determined $J$ of Hydrus~1.

The relative $J$ values among different dwarfs are an essential ingredient in joint searches for an annihilation signal from all dwarfs at once {\citep[e.g.][]{2011PhRvL.107x1303G,2011PhRvL.107x1302A,2012APh....37...26M}}. Pointed instruments (e.g. imaging atmospheric Cherenkov telescopes) must incorporate the $J$ rankings in designing observing strategies. Ratios of $J$'s also provide a crucial test of a dark matter origin for any signal detected from a dwarf, since fluxes must be proportional to $J$~\citep{2007PhRvD..75h3526S,geringer15,2015PhRvL.115h1101G,2017ApJ...834..110A}. Although Hydrus~1 is not among the top tier of dwarfs in terms of expected annihilation flux it outranks most classical and ultrafaint dwarfs that are commonly used to constrain dark matter particle physics.

\begin{figure}
	\includegraphics[width=\columnwidth]{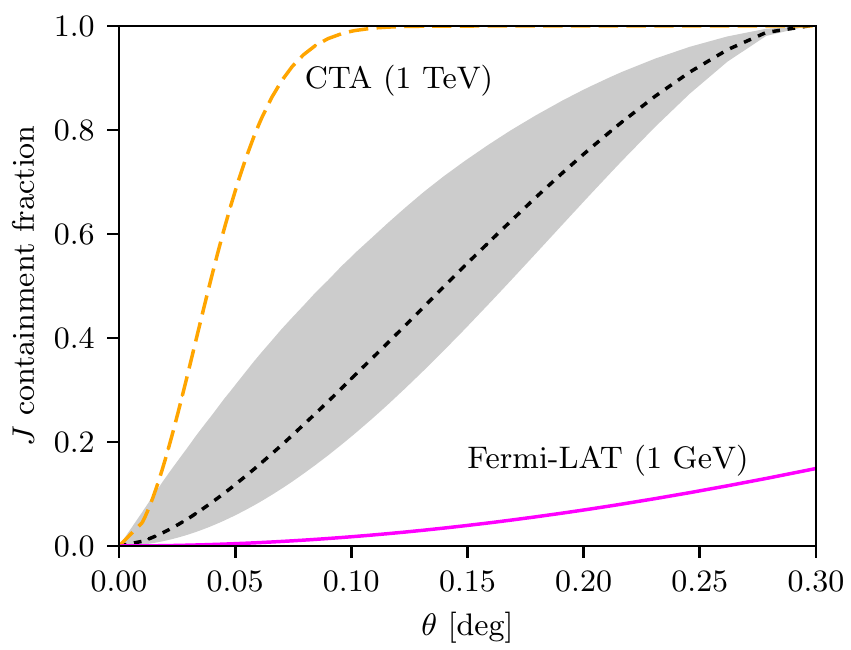}
    \caption{Constraints on the spatial extent of Hydrus~1's dark matter annihilation profile. For each angle $\theta$ away from the center of the dwarf the $J$ containment fraction gives the fraction of annihilation flux that comes from the region within $\theta$. The black dotted line is the median estimate, while the gray band shows the 16th to 84th percentiles of the posterior. The gamma-ray PSFs of the CTA (at 1~TeV) and Fermi-LAT (at 1~GeV) are shown for comparison. CTA will detect Hydrus~1 as an extended source.}
    \label{fig:Jcdf}
\end{figure}

Hydrus~1 is close enough to earth that its dark matter annihilation may have a detectable spatial extent. A detection in gamma-rays would be a direct observation of the shape of Hydrus~1's dark matter density profile. Following~\citet{geringer15}, Fig.~\ref{fig:Jcdf} illustrates Hydrus~1's spatial extent. For each angle $\theta$, the shaded region shows a 68\% credible interval on the fraction of annihilation flux coming from angles less than $\theta$. Approximate PSFs for Fermi-LAT and CTA are shown for comparison\footnote{They are modeled as 2-d Gaussian distributions with 68\% containment fractions of $0.05\degr$ for CTA at 1~TeV and $0.8\degr$ for Fermi-LAT at 1~GeV.}. If CTA is able to detect Hydrus~1 in the future it will almost certainly be seen as an extended source. The dark matter halo truncation radius we adopt is much smaller than the scale of the Fermi-LAT PSF. If Hydrus~1's halo extends well beyond the outermost spectroscopically-observed member star it may appear extended to Fermi. However, this question is beyond the scope of our conservative analysis.

Finally, we perform a search for gamma-ray emission from Hydrus~1 using data from the Fermi Large Area Telescope~\citep{2009ApJ...697.1071A}. Hydrus is located $0.45\degr$ in projection from a gamma-ray source {\tt 3FGL J0224.1-7941} in the Third Fermi Source Catalog (3FGL)~\citep{2015ApJS..218...23A}, which is associated with a probable blazar. This precludes performing the search using optimized event-weighting~\citep[e.g.][]{2015PhRvD..91h3535G,2015PhRvL.115h1101G} as that method does not take into account contamination from the bright nearby source. We therefore adopt the commonly used approach based on maximum likelihood modeling of the region surrounding Hydrus~1. A best-fitting model is found both with and without a trial source at the location of Hydrus~1. If the likelihood is sufficiently improved by the addition of the trial source it is interpreted as evidence for emission from the direction of Hydrus~1.

Gamma-ray events from a $10\degr \times 10\degr$ region centered on Hydrus~1 are binned into $0.05\degr$ spatial pixels and 25 logarithmic energy bins from 1~GeV to 316~GeV (10 bins per decade). The spatial binning is substantially smaller than the Fermi PSF at all energies above 1~GeV and the analysis will be as sensitive as if we had used an unbinned likelihood function. Restricting the energy range to be greater than 1~GeV avoids events with poor direction reconstruction and does not negatively impact the power of the search for dark matter for particle masses greater than a few GeV~\citep[Sec. VI D]{2015PhRvD..91h3535G}. Our data set comprises 9.4 years of observation\footnote{We use version v10r0p5 of the Fermi Science Tools. Front and back \texttt{P8R2\_SOURCE\_V6} events from weeks 9 to 500 were extracted with \texttt{gtselect} (\texttt{zmax=90}), and \texttt{gtmktime} (\texttt{filter="(DATA\_QUAL>0) \&\& (LAT\_CONFIG==1)"} and \texttt{roicut=no}). We ran \texttt{gtltcube} with \texttt{zmax=90} and \texttt{gtexpcube2} with \texttt{binsz=0.1} to create an exposure map, \texttt{gtpsf} to compute the PSF at the location of Hydrus~1, and assume that the PSF is constant across region surrounding Hydrus~1.}.
The likelihood function is the product of an independent poisson probability for each bin. The predicted number of events in a bin is the sum of isotropic\footnote{\texttt{iso\_P8R2\_SOURCE\_V6\_v06.txt}} and diffuse\footnote{\texttt{gll\_iem\_v06.fits}} background components\footnote{At the edges of each energy bin the background templates were multiplied by the exposure map, convolved with the PSF at a resolution of $0.05\degr$ over a $25\degr \times 25\degr$ region, and integrated over the energy bin (assuming log flux $\propto$ log energy) to obtain the model prediction.}
along with the four 3FGL sources within the $10\degr \times 10\degr$ region. These sources have all been associated with extragalactic objects by the Fermi Collaboration in the 3FGL and we adopt the precise locations of their counterparts rather than their coordinates in the 3FGL. Each source is modeled as a point source with a ``log-parabola'' energy spectrum, whose flux takes the form $dF/dE \propto E^{-(\alpha + \beta \log E)}$. The null model has 14 free parameters: normalizations of the two background components and the normalization, $\alpha$, and $\beta$ of each source. The trial point source located at the position of Hydrus~1 has an energy spectrum given by $dF/dE \propto E^{-(\alpha + \beta \log E)} \exp(-E/E_c)$, with 4 additional free parameters (normalization, $\alpha$, $\beta$, and $E_c$). This flexible energy spectrum is sufficient to describe dark matter annihilation into Standard Model leptons, quarks, and gauge bosons.

The test statistic is $\lambda_\mathrm{obs} = 2\log (\hat{L}_1/ \hat{L}_0)$, where $\hat{L}_0$ is the value of the likelihood maximized over the 14 free parameters of the null model without Hydrus~1 and $\hat{L}_1$ is the maximum likelihood value when all 18 parameters are free to vary. The detection significance ($p$ value) is the probability that, if the null model were true, $\lambda$ would be measured to be larger than $\lambda_\mathrm{obs}$. 
After separately maximizing over the 14 and 18 parameter model we measure $\lambda_\mathrm{obs}=1.73$. Calibrating the distribution of $\lambda$ using the region surrounding Hydrus~1, we find the probability of obtaining a larger $\lambda$ under the null model is $\gtrsim 21\%$ and conclude that the Fermi data do not show any evidence for gamma-ray emission from Hydrus~1. This is consistent with gamma-ray observations of dwarfs with larger $J$ values~\citep[e.g.][]{2017ApJ...834..110A}, including both non-detections as well as a potential gamma-ray signal from Reticulum~II if the latter is due to dark matter annihilation~\citep{2015PhRvL.115h1101G,2015JCAP...09..016H} (see also~\citet{2015ApJ...809L...4D,2017ApJ...834..110A}). We also search for new sources in the vicinity of Hydrus~1 by adding a trial source at each location within the $10\degr \times 10\degr$ region. This search reveals two probable new gamma-ray sources but they are several degrees away from Hydrus~1.

\subsection{Chemistry}
\label{sec:chemistry_discussion}

\begin{figure}
	\includegraphics[width=0.5\textwidth]{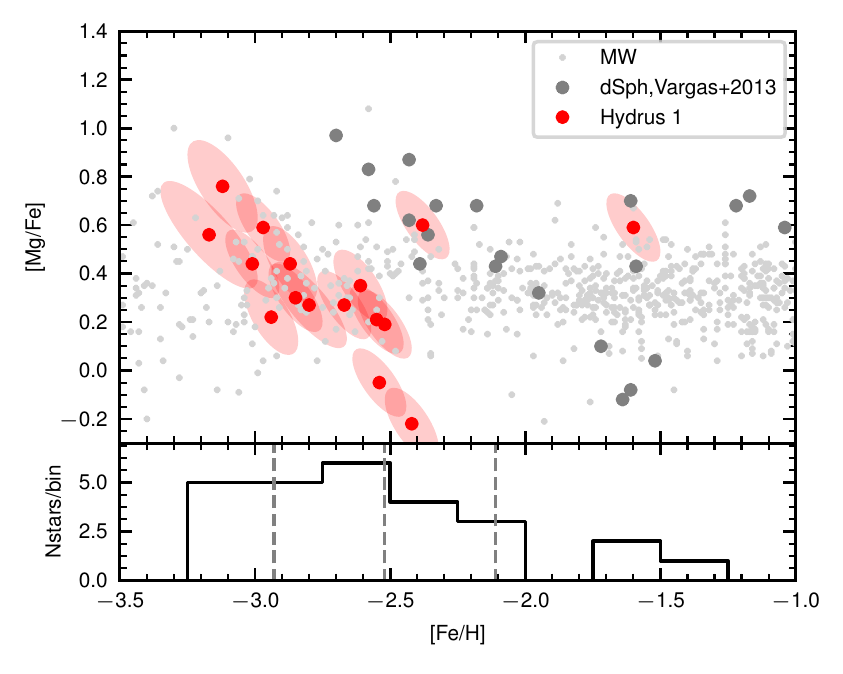}
	\caption{Chemical abundances of Hydrus~1 member stars. {\it Top panel:} [Mg/Fe] vs [Fe/H] for Hyi~1 stars are shown by red points with ellipses showing the 1 sigma covariance. Milky Way stars  from the SAGA database \citep{suda08} are shown by small light grey markers  while  stars from several faint dwarf spheroidals \citep{vargas13} are shown by large dark grey circles. One Hydrus~1 star (star 26) is outside the plot, with [Fe/H]$\sim -2.15$ and [Mg/Fe]$\sim-0.9$ {\it Bottom panel:} Histogram of iron abundances of Hydrus~1 stars. The dashed lines show the mean and standard deviation of the Gaussian abundance distribution that was part of our chemo-dynamical model. We note that the metallicity distribution seems to be non-Gaussian, with a more pronounced metal-rich tail.}  
\label{fig:mgfe}
\end{figure}

In Section~\ref{sec:chemistry} we described that we were able to get robust abundances of more than one chemical element, in particular iron, magnesium and carbon for many Hydrus members. First we look at the abundances of alpha elements vs iron elements in the satellite. The alpha element abundances are particularly useful for constraining the length of star formation history, due to the fact that Type Ia and Type II Supernovae have vastly different yields in alpha-elements and that Type Ia supernovae occur with a delay of $\sim$ 100\,Myr after the beginning of star formation and the first Type II supernovae explosions \citep{woosley95,maoz14}.  
Figure~\ref{fig:mgfe} shows the [Mg/Fe] vs [Fe/H] abundance distribution for Hydrus~1 member stars, together with the distribution in the MW halo and other UFDs. We note that, on average, the [Mg/Fe] in Hydrus's stars is larger than zero $<$[Mg/Fe]$>\sim 0.2$, which is indicative that the star-formation episode that led to the formation of Hydrus~1 was not very extended in time. However, the data also show that [Mg/Fe] decreases significantly from [Mg/Fe]$\sim 0.6$ at $[Fe/H]\sim -3$ to almost solar at higher metallicity. This feature is associated with self-enrichment of gas  by Type Ia supernovae \citep{tinsley79} and implies that the star formation proceeded for at least $\sim$ 100\, Myr so that the $\alpha$ abundances were diluted by Type Ia yields. The lack of higher metallicity stars with solar alpha-to-iron abundances, however, means that the star formation did not proceed longer than $100$\,Myr-$1$\,Gyr. A similar decline in alpha abundances is very pronounced in the MW halo and MW thick disk, as well as in some classical dwarfs \citep{tolstoy09} and more luminous UFDs \citep{vargas13}, where it tends to occur at higher metallicities. In Hydrus~1 this knee occurs at a metallicity of $-2.6$. 

On the bottom panel of Figure~\ref{fig:mgfe} we also show the metallicity distribution function (MDF) from spectral synthesis modeling of Hyi~1 stars. The distribution seems to be asymmetric with a longer tail towards high metallicities, thereby challenging our assumption of Gaussian metallicity distribution for the purposes of our chemo-dynamical modelling.  However, because of small number statistics and significant uncertainties in individual metallicity measurements, it remains to be seen whether the MDF asymmetry is real.  We also remark that most dwarf galaxies show MDFs with a more pronounced metal-poor tail; however, some, such as Ursa Minor \citep{kirby11}, show a metal rich tail, which could be explained by chemical evolution models with lower galactic wind efficiency.

Two likely Hyi~1 member stars with higher metallicities shown in Fig.~\ref{fig:mgfe} have high [Mg/Fe] ratio. The cause of this is unclear. It is possible, although unlikely, that those are MW halo contaminants rather than Hydrus~1 stars. We also remark on one star (star 26) that is outside the plotting range on Figure~\ref{fig:mgfe} with unusually low [Mg/Fe]$\sim-0.9$. Similar abundances were previously observed in some chemically peculiar stars in other dwarfs, such as Carina \citep{venn12}, where the unusual abundance patters were potentially associated with inhomogeneous gas mixing and star formation from SN Ia enriched gas,  leading to an overabundance of iron with respect to other elements.

Upper limits on [C/H] are available for 21 Hydrus~1 members. These limits are not informative. We can only confirm that one of the more metal-rich stars, Star 77 with $[{\rm Fe}/{\rm H}] = -1.60 \pm 0.10$, is not a carbon-enhanced metal-poor star: $[{\rm C}/{\rm Fe}] < 0.54$. For all other members with upper limits on [C/H], the limits are higher than both the $[{\rm C}/{\rm Fe}] > +0.7$ and $[{\rm C}/{\rm Fe}] > +1.0$ definitions that are commonly adopted for carbon-enhanced metal-poor stars \citep{Aoki_2007,Beers_Christlieb_2005}. From these limits we cannot summarise how many Hydrus~1 members meet the definition for carbon enhancement.

With $[{\rm C}/{\rm Fe}] \sim +3$ and $[{\rm Fe}/{\rm H}] = -3.18$, Star 75 clearly meets the definition of a carbon-enhanced metal-poor star. Indeed, the estimated [C/H] ratio is near the Solar value, which is remarkable for an extremely metal-poor star in an ultra-faint dwarf galaxy. A search of the SAGA database \citep{suda08} and the JINAbase \citep{jinabase} reveals that Star 75 of Hydrus~1 is the most carbon-enhanced metal-poor star known in a dwarf galaxy. The next most carbon-enhanced stars known in dwarf galaxies are Star 11 in Bo\"otes I \citep{Lai_2011}, and Star 7 in Segue 1 \citep{Frebel_2014}, both which have $[{\rm C}/{\rm Fe}] \approx +2.3$. Similarly, in a study of 67 carbon-enhanced metal-poor stars in the Milky Way halo, only one was found to have $[{\rm C}/{\rm Fe}] > 3$ \citep{Abate_2015}.

Strong carbon enhancement in metal-poor stars results in depressed flux levels at bluer wavelengths (making $g$ relatively fainter than $r$), causing a reddening in the $g-r$ colour. This is consistent with what we observe in the second panel of Figure \ref{fig:spec_plot}. Star 75, as marked with a cross-hair, is some $\sim$0.1 magnitudes redder than most other members. This effect has been discussed before \citep{Aoki_2002}.  Since candidate members in ultra-faint dwarf galaxies are usually selected for follow-up spectroscopy based on their proximity to an isochrone (among other factors), it is likely that any similarly carbon-enhanced metal-poor stars in other ultra-faint dwarf galaxies have been overlooked. This acts to further complicate any estimates of the carbon-enhanced metal-poor star fraction among these near pristine galaxies \citep{Frebel_2015}.

\section{Conclusions}
\label{sec:conclusions}

In this paper we presented the discovery and detailed analysis of
a new ultra-faint Milky Way satellite, called Hydrus 1. Below we
summarize the key findings about the system.

\begin{itemize} 
\item The satellite is located in between the Large and the Small
  Magellanic Clouds on the sky at the heliocentric distance of
  $28$\,kpc and $24$\,kpc from the LMC. The luminosity of the
  satellite is $M_V\sim -4.7$ and the color-magnitude diagram
  indicates an old, metal-poor stellar population with several BHB
  stars and two RR-Lyrae.  The object is mildly elliptical with an
  axis ratio of $0.8$, and is approximately aligned with the right
  ascension axis and proper motion of LMC. The circularized half-light
  radius of Hydrus~1 is $50$\,pc, and it has a surface brightness of $\sim
  27.5$\,mag/sq. arcsec, which is noticeably brighter than many typical
  ultra-faint dwarf spheroidals.  These properties place the object in an ambiguous
  region between extended globular clusters and dwarf galaxies in the plane of size and luminosity.

\item High-resolution spectroscopy revealed
  $\sim$ 30 confirmed members of Hydrus~1 with mean $V_{GSR}$ of
  $-94$\,km\,s$^{-1}$ and a well resolved stellar velocity dispersion of
  $2.7$\,km\,s$^{-1}$. The implied mass-to-light ratio within the
  half-light radius is $\log_{10} M/L =1.82\pm 0.16$, thus indicating 
  that the system is dark matter dominated. Nevertheless, the mass-to-light ratio is uncharacteristically low in comparison to other dwarfs
  of similar luminosity. We speculate that it is possible that many
  other Galactic faint satellites with ambiguous sizes between
  extended clusters and dwarf galaxies also have intermediate values
  of mass-to-light ratios. These systems may either be globular
  clusters with a small amount dark matter or dwarf galaxies that lost
  some dark matter due to tides. Our analysis of the stellar
  kinematics of Hydrus~1 also reveals a tentative (at 2.5 sigma
  significance level) line-of-sight velocity gradient, which is compatible
  with a $3$\,km\,s$^{-1}$ rotation of the satellite. If real, this
  could signify the detection of a first rotating ultra-faint dwarf
  galaxy.

\item Based on the spatial position and the observed line-of-sight velocity, 
  we find it highly plausible that Hydrus~1 is part of the Magellanic
  family of satellites. The dwarf may currently be a part of the
  leading debris of the LMC, having been stripped from the LMC between
  0.1\,Gyr and 0.5\,Gyr ago. A connection between Hydrus~1 and the
  LMC should be confirmed or refuted by the incoming Gaia DR2 data.

\item Due to its relative proximity to the Sun, Hydrus 1 is a very
  promising target for dark-matter annihilation searches, with a
  J-factor of $\log J=18.33\pm0.36$.  While it is not the highest
  J-factor among all the MW dwarfs, it is competitive and has a 
  small uncertainty, thanks to the high quality velocity dispersion
  measurement. Our analysis of the Fermi data at the location of
  Hydrus for the dark matter annihilation reveals no
  significant signal.

\item Chemically, Hydrus~1 is rather metal poor, with mean [Fe/H] $\sim
  -2.5$ and significant scatter in abundances of $0.4$\,dex. It has an
  enhanced alpha-abundance <[Mg/Fe]>$\sim 0.2$ and shows evidence of
  self-enrichment by SN Ia at timescales of 100-1000\,Myr demonstrated
  by the decreasing [Mg/Fe] with increasing [Fe/H]. Analysis of 30
  Hydrus~1 member stars revealed one carbon-enhanced extremely
  metal-poor star with metallicity [Fe/H]<$-3$ and [C/Fe] $\sim
  3$. This level of carbon enrichment makes it one the most
  carbon-enhanced stars known in both the MW halo and the dwarf
  galaxies.
\end{itemize}

\section*{Acknowledgements}

A.~R.~C. acknowledges support from the
Australian Research Council (ARC) through Discovery Project
grant DP160100637. D.M. is supported by an ARC Future Fellowship (FT160100206).  D.M. and G.D.C. acknowledge support from ARC Discovery Project DP150103294. M.M. acknowledges support from NSF grant AST1312997. M.G.W. acknowledges support from NSF grant AST1412999. E.O. acknowledges support from NSF grant AST1313006. The research leading to these results has received funding from the European Research Council under the European Union's Seventh Framework Programme (FP/2007-2013) / ERC Grant Agreement n. 308024.

% standard decam acq https://www.noao.edu/noao/library/NOAO_Publications_Acknowledgments.html
This project used data obtained with the Dark Energy Camera (DECam),
which was constructed by the Dark Energy Survey (DES) collaboration.
Funding for the DES Projects has been provided by 
the U.S. Department of Energy, 
the U.S. National Science Foundation, 
the Ministry of Science and Education of Spain, 
the Science and Technology Facilities Council of the United Kingdom, 
the Higher Education Funding Council for England, 
the National Center for Supercomputing Applications at the University of Illinois at Urbana-Champaign, 
the Kavli Institute of Cosmological Physics at the University of Chicago, 
the Center for Cosmology and Astro-Particle Physics at the Ohio State University, 
the Mitchell Institute for Fundamental Physics and Astronomy at Texas A\&M University, 
Financiadora de Estudos e Projetos, Funda{\c c}{\~a}o Carlos Chagas Filho de Amparo {\`a} Pesquisa do Estado do Rio de Janeiro, 
Conselho Nacional de Desenvolvimento Cient{\'i}fico e Tecnol{\'o}gico and the Minist{\'e}rio da Ci{\^e}ncia, Tecnologia e Inovac{\~a}o, 
the Deutsche Forschungsgemeinschaft, 
and the Collaborating Institutions in the Dark Energy Survey. 
The Collaborating Institutions are 
Argonne National Laboratory, 
the University of California at Santa Cruz, 
the University of Cambridge, 
Centro de Investigaciones En{\'e}rgeticas, Medioambientales y Tecnol{\'o}gicas-Madrid, 
the University of Chicago, 
University College London, 
the DES-Brazil Consortium, 
the University of Edinburgh, 
the Eidgen{\"o}ssische Technische Hoch\-schule (ETH) Z{\"u}rich, 
Fermi National Accelerator Laboratory, 
the University of Illinois at Urbana-Champaign, 
the Institut de Ci{\`e}ncies de l'Espai (IEEC/CSIC), 
the Institut de F{\'i}sica d'Altes Energies, 
Lawrence Berkeley National Laboratory, 
the Ludwig-Maximilians Universit{\"a}t M{\"u}nchen and the associated Excellence Cluster Universe, 
the University of Michigan, 
{the} National Optical Astronomy Observatory, 
the University of Nottingham, 
the Ohio State University, 
the OzDES Membership Consortium
the University of Pennsylvania, 
the University of Portsmouth, 
SLAC National Accelerator Laboratory, 
Stanford University, 
the University of Sussex, 
and Texas A\&M University.

This work has made use of data from the European Space Agency (ESA)
mission {\it Gaia} (\url{https://www.cosmos.esa.int/gaia}), processed by
the {\it Gaia} Data Processing and Analysis Consortium (DPAC,
\url{https://www.cosmos.esa.int/web/gaia/dpac/consortium}). Funding
for the DPAC has been provided by national institutions, in particular
the institutions participating in the {\it Gaia} Multilateral Agreement.

Based on observations at Cerro Tololo Inter-American Observatory,
National Optical Astronomy Observatory (NOAO Prop. 2016A-0618 and 2017B-0906; 
PI: D. Mackey), which is operated
by the Association of Universities for Research in Astronomy (AURA)
under a cooperative agreement with the National Science
Foundation.

% Don't change these lines
\bsp	% typesetting comment
\label{lastpage}
\end{document}